\documentclass[aps,prd,preprint,titlepage,superscriptaddress,longbibliography,amsmath,amssymb]{revtex4-2}

\usepackage{bm}
\usepackage{xcolor}
\usepackage{tikz}          % inline Feynman diagrams

\usetikzlibrary{calc,decorations.markings,decorations.pathmorphing,
                arrows.meta,positioning,fit,backgrounds,
                patterns,shapes.geometric}

\providecommand{\SGfill}{black!8}     % subgraph shading
\providecommand{\Blobfill}{black!18}  % dressed blob fill
\providecommand{\HLcol}{black!22}     % subgraph-highlight halo

\tikzset{
  prop/.style   = {line width=0.7pt},
  extl/.style   = {line width=0.7pt},
  dprop/.style  = {line width=0.7pt,
                   postaction={decorate,
                     decoration={markings,
                       mark=at position 0.5 with {
                         \fill[\Blobfill,draw=black,line width=0.4pt]
                           circle[radius=1.5pt];}}}},
  vtx/.style    = {circle,fill=black,inner sep=0pt,minimum size=5pt},
  blob/.style   = {circle,draw=black,fill=\Blobfill,line width=0.5pt,
                   inner sep=0pt,minimum size=9pt},
  subg/.style   = {draw=black,dashed,line width=0.6pt,rounded corners=3pt,
                   fill=\SGfill,inner sep=3pt},
  hl/.style     = {line width=4pt,\HLcol,line cap=round},
}

\newcommand{\extleg}[4]{%
  \draw[extl] #1 -- #2;%
  \ifx\relax#3\relax\else\node[font=\scriptsize,#4] at #2 {#3};\fi}

\tikzset{sqmark/.style={circle,draw=black,fill=white,line width=0.5pt,
                        inner sep=0pt,minimum size=3.4pt}}
\tikzset{
  gluon/.style = {decorate,line width=0.6pt,
                  decoration={coil,aspect=0.55,segment length=2.1mm,
                              amplitude=0.9mm}},
  ghost/.style = {line width=0.7pt,dash pattern=on 2.1pt off 1.7pt},
  quark/.style = {line width=0.8pt},
  ggvtx/.style = {circle,fill=black,inner sep=0pt,minimum size=5pt},
  ggarr/.style = {postaction={decorate,decoration={markings,
                   mark=at position 0.57 with {\arrow{Stealth[length=1.6mm]}}}}},
}

\usepackage[T2A]{fontenc}\usepackage[utf8]{inputenc}
\usepackage[english]{babel}
\usepackage{graphicx}
\usepackage{hyperref}

\usepackage{bm}
\usepackage{mathbbol}
\usepackage{amsthm}
\usepackage{mathrsfs}
\usepackage{latexsym}
\usepackage{commath} %\abs is defined here

\usepackage{graphicx}
\usepackage{tikz}
\usetikzlibrary{babel,arrows.meta,positioning}
\usepackage{appendix}

\usepackage{subfig}
\usepackage{subcaption}
\usepackage{xcolor}

\graphicspath{{images/}}

\newcommand{\be}{\begin{equation}}
\newcommand{\ee}{\end{equation}}

\newcommand{\benn}{\begin{equation*}}
\newcommand{\eenn}{\end{equation*}}

\newcommand{\bea}{\begin{equation}\begin{aligned}}
\newcommand{\eea}{\end{aligned}\end{equation}}

\usepackage{mathtools}

\DeclarePairedDelimiterX\braket[2]{\langle}{\rangle}{#1 \delimsize\vert #2}

\usepackage{physics}
\usepackage{tensor}
\usepackage{graphicx}
\usepackage{xcolor}
\usepackage{cleveref}
\usepackage{lipsum}
\usepackage{comment}
\usepackage{siunitx} %Uncomment this line if it is needed
\begin{document}

\begin{flushright}
   INR-TH-2026-017
\end{flushright}

\title{Attractors and initial-condition sensitivity in Horndeski cosmologies}

%\author[a,b,c,1]{Y.~A.~Ageeva }
%\author[a,c,2]{and A.~K.~Sharofeev}

%\affiliation[a]{Institute for Nuclear Research of %the Russian Academy of
%  Sciences, 60th October Anniversary Prospect, %7a, 117312 Moscow, Russia}
%\affiliation[b]{Institute for Theoretical and %Mathematical Physics,
%  M.V. Lomonosov Moscow State University, %Leninskie Gory 1, 119991
%  Moscow, Russia}
%\affiliation[c]{Department of Particle Physics and Cosmology, Physics Faculty,
%  M.V. Lomonosov Moscow State University, %Leninskie Gory 1-2, 119991
%  Moscow, Russia}
%\emailAdd{$^{1}$ageeva@inr.ac.ru, %$^{2}$sharofeev@inr.ac.ru}

\author{Y.~A.~Ageeva}
\email{ageeva@inr.ac.ru}
\affiliation{Institute for Nuclear Research of the Russian Academy of
  Sciences, 60th October Anniversary Prospect, 7a, 117312 Moscow, Russia}
\affiliation{Institute for Theoretical and Mathematical Physics,
  M.V. Lomonosov Moscow State University, Leninskie Gory 1, 119991
  Moscow, Russia}
\affiliation{Department of Particle Physics and Cosmology, Physics Faculty,
  M.V. Lomonosov Moscow State University, Leninskie Gory 1-2, 119991
  Moscow, Russia}

\author{A.~K.~Sharofeev}
\email{sharofeev@inr.ac.ru}
\affiliation{Institute for Nuclear Research of the Russian Academy of
  Sciences, 60th October Anniversary Prospect, 7a, 117312 Moscow, Russia}
\affiliation{Department of Particle Physics and Cosmology, Physics Faculty,
  M.V. Lomonosov Moscow State University, Leninskie Gory 1-2, 119991
  Moscow, Russia}

\begin{abstract}
We study whether nearby homogeneous cosmological histories approach or move away from the Genesis and contracting solutions. We also determine how an uncertainty in the scalar velocity at a later stage affects the inferred initial velocity. In the Genesis models, including the gravitational dynamics can substantially change this relation. We find that convergence of homogeneous solutions can occur even around backgrounds with unstable perturbations, and therefore attraction of the background is not sufficient to establish that a cosmological solution is physically viable. In the contracting model, one power-law solution attracts nearby trajectories, whereas another repels them. The strength of these effects depends on how the cosmological state is parametrized. Their evolution through the subsequent bounce is not considered here.
\end{abstract}

\maketitle

%%%%%%%%%%%%%%%%%%%%%%
\section{Introduction}
\label{sec:Introduction}
%%%%%%%%%%%%%%%%%%%%%%

Inflation provides a successful explanation of the large-scale homogeneity and spatial flatness of the Universe~\cite{Guth:1980zm,Linde:1981mu} and a mechanism for generating primordial perturbations~\cite{Mukhanov:1981xt,Starobinsky:1982ee}. Its success nevertheless leaves open the question of the preceding cosmological evolution: an inflationary stage does not by itself supply a nonsingular description of the past~\cite{Borde:2001nh}. This motivates the study of alternative early-universe histories that can generate primordial fluctuations and of nonsingular stages that may precede inflation. Establishing whether such histories can be realized consistently also clarifies which predictions depend specifically on inflation. For further classic formulations of inflation and the generation of primordial perturbations, see~\cite{Starobinsky1980,Sato1981,AlbrechtSteinhardt1982,Hawking1982,GuthPi1982,Starobinsky:1982ee,LindeFluctuations1982,Linde1983,BardeenSteinhardtTurner1983,MukhanovFeldmanBrandenberger1992,GarrigaMukhanov1999,Chibisov:1982nx,Mukhanov:2005sc,Gorbunov:2011zzc}.

A central difficulty in constructing nonsingular cosmologies is obtaining a stage with an increasing Hubble parameter. In general relativity, for a spatially flat homogeneous and isotropic Universe sourced by minimally coupled matter, $\dot{H} > 0$ requires violation of the null energy condition. Such violation is difficult to reconcile with stable perturbations in simple scalar-field theories~\cite{Rubakov:2014jja}. Horndeski theory provides a broader framework, incorporating derivative interactions and scalar--metric couplings while retaining second-order field equations~\cite{Horndeski:1974wa,Kobayashi:2011nu} (recent higher-derivative extensions include covariant
scalar--tensor and scalar--vector--tensor
constructions~\cite{Gorji:2026CovariantST,Gorji:2026HOMES},
as well as scalar--tensor theories with up to third-order
derivatives that satisfy degeneracy and consistency conditions
ensuring the propagation of one scalar and two tensor degrees of
freedom and extend generalized disformal Horndeski and U-DHOST
theories~\cite{Michiwaki:2026Healthy}). These interactions allow backgrounds with $\dot{H} > 0$ and stable linear perturbations over suitable intervals, as illustrated by Galilean Genesis~\cite{Creminelli:2010ba}. For nonminimally coupled models, the geometric condition $\dot{H} > 0$ must be distinguished from an energy condition imposed on a separately defined matter component~\cite{Kobayashi:2019hrl}.

Two realizations of this idea are Galilean Genesis and nonsingular bouncing cosmology. In Genesis, the Universe expands from an asymptotically Minkowski state: the Hubble parameter approaches zero in the asymptotic past and increases during the Genesis stage. A preceding contracting phase is not required~\cite{Creminelli:2010ba}. Generalized Galilean Genesis extends this construction to families with different background scalings and primordial spectra~\cite{Nishi:2015pta}. Further realizations, extensions, and consistency analyses of Genesis cosmologies have been presented in~\cite{Creminelli:2012my,Hinterbichler:2012fr,Elder:2013gya,Pirtskhalava:2014esa,Kobayashi:2015gga,Kolevatov:2017voe,Mironov:2019haz,Mironov:2019qjt,Ageeva:2020gti,Volkova:2026Cuscuton,Choi:2025Genesis,Choi:2025GenesisStarobinsky}. Moreover, recently a geodesically complete to the past
Genesis phase, with
NEC-respecting matter and a nearly scale-invariant scalar spectrum,
was constructed in type-II minimally modified gravity. Its asymptotic
cubic power counting shows no strong-coupling scale collapsing to zero
at fixed nonzero momenta \cite{Ganz:2026mek}. In the specific ``Subluminal Galilean Genesis'' model of
Ref.~\cite{Creminelli:2012my}, scalar perturbations on the Genesis
background are substantially subluminal, and this property is robust
under sufficiently small deformations of that background.
Ref.~\cite{Easson:2013WhenMatter} further showed that, in the absence
of additional matter, scalar perturbations are subluminal throughout
the spatially flat cosmological phase space, whereas adding matter
coupled to the Galileon only through gravity opens regions of phase
space with superluminal scalar propagation. A bounce instead connects contraction to expansion through a positive minimum of the scale factor. Within kinetic gravity braiding, Ref.~\cite{Easson:2011zy}
constructed a broad class of spatially flat bouncing solutions that
are free of ghost and gradient instabilities around the turning point.
For a reconstructed bounce in the same class,
Ref.~\cite{Dobre:2017Unbraiding} performed a phase-space analysis and
found additional stable bouncing trajectories together with stable
superluminal configurations continuously connected to the selected
bounce. In suitable Horndeski models, primordial perturbations can be generated during a power-law contracting stage preceding the bounce~\cite{Ageeva:2022asq}. For reviews and representative constructions of nonsingular bouncing cosmologies, see~\cite{Novello:2008ra,Lehners:2008vx,Lehners:2011kr,Battefeld:2014uga,Qiu:2011cy,Easson:2011zy,Cai:2012va,Osipov:2013ssa,Qiu:2013eoa,Koehn:2013upa,Battarra:2014tga,Qiu:2015nha,Ijjas:2016tpn,Brandenberger:2016vhg,Ijjas:2018qbo,Kolevatov:2016ppi,Cai:2017dyi,Kolevatov:2017voe,Mironov:2018Bounce,Ye:2019sth,Ilyas:2020qja,Mironov:2020Subluminal,Zhu:2021whu,Mironov:2022Disformal}.

The existence of a background solution is only the first step toward a viable cosmological scenario. Scalar and tensor perturbations must be free from ghost and gradient instabilities, and these conditions must remain satisfied through the transitions connecting successive stages~\cite{Kobayashi:2011nu}. Superluminal scalar propagation is distinct from ghost and gradient
instabilities and does not by itself establish a classical
inconsistency of a cosmological background; it can instead be relevant
to the possibility of a standard local, Lorentz-invariant Wilsonian
ultraviolet completion, which must be assessed over the configurations
belonging to the same connected effective-theory domain rather than on
a single selected trajectory
~\cite{Easson:2013WhenMatter,Dobre:2017Unbraiding}. No-go theorems constrain the realization of globally nonsingular and linearly stable bouncing and Genesis histories within spatially flat Horndeski cosmology under specified assumptions~\cite{Libanov:2016kfc,Kobayashi:2016xpl}. Moreover, stability at quadratic order does not ensure perturbative control: the relevant strong-coupling scales must remain well above the characteristic energies of the background evolution and of the perturbation modes under consideration. This requirement imposes additional restrictions on both Genesis and contracting models~\cite{Akama:2022usl,Ageeva:2022asq}. Related analyses of classical consistency, strong coupling, perturbative unitarity, and gravitational-wave production in early-Universe models were carried out in~\cite{Ageeva:2020Classical,Ageeva:2021StrongGravity,Ageeva:2022StrongCoupling,Ageeva:2023Unitarity,Ageeva:2024KInflation,Petrov:2026GWUnitarity}.

Observational viability requires a scalar power spectrum consistent with the measured amplitude and spectral tilt~\cite{Planck:2018vyg,AtacamaCosmologyTelescope:2025blo}, a tensor contribution compatible with observational bounds~\cite{BICEP:2021xfz}, and primordial non-Gaussianity within the corresponding limits~\cite{Planck:2019kim}. Power spectra have been studied for generalized Genesis~\cite{Nishi:2015pta} and for the contracting Horndeski model considered here~\cite{Ageeva:2022asq}. A nearly scale-invariant scalar spectrum alone, however, does not establish viability: non-Gaussianity and weak coupling impose additional consistency requirements, as investigated for Genesis in~\cite{Akama:2022usl} and for the contracting model in~\cite{Ageeva:2024knc}. At exact scale invariance, Ref.~\cite{Akama:2025FullBispectra}
computed the complete scalar and tensor auto- and cross-bispectra for
a general class of power-law contracting backgrounds in Horndeski
theory that encompasses the corresponding limit of the contracting
construction considered here. That analysis distinguishes a branch
with conserved superhorizon modes, for which the full set of
bispectra obeys Maldacena's consistency relation, from a growing-mode
branch in which this relation is violated. Relating these early-time predictions to observations also requires control over the evolution of perturbations through subsequent cosmological transitions.

A separate question concerns the dependence of the background evolution on its initial data. An explicitly constructed solution may attract or repel neighboring homogeneous trajectories, and even an attracting solution may have a restricted basin of attraction. Attractor behavior has been demonstrated for generalized Genesis backgrounds~\cite{Nishi:2015pta}. The existence of an attractor alone, however, does not specify how strongly a chosen final quantity depends on initial data over a finite evolution interval. Suppression of this dependence can also make the inverse problem sensitive to uncertainties: small errors in the final data may correspond to substantially larger errors in the reconstructed initial values. Quantifying this effect requires a specified choice of variables and initial and final sections of the evolution. Homogeneous attraction must also be distinguished from perturbative stability, since convergence of background trajectories does not establish the absence of ghost and gradient instabilities.

In this work, we investigate the dependence on initial data for two generalized Genesis solutions and the power-law contracting stage of a Horndeski bouncing model. Using a common reduction of the background equations, we analyze their homogeneous phase portraits for each regular real value of the Hubble parameter allowed by the background equation. With variables $\psi = e^{-\lambda \phi}$ and $v= \dot{\psi}$, we quantify the local dependence of the final velocity on its initial value between fixed scalar-field sections, together with the conditioning of the inverse reconstruction. For the Genesis models, we compare the scalar-field approximation with the full background dynamics to assess the effect of gravitational backreaction. The resulting diagnostic is defined for the chosen velocity variable, comparison sections, and continuously followed Hubble solution. For the contracting model, our analysis covers only the pre-bounce stage; extending it through the interpolating bounce remains a separate problem.

For the two Genesis models, we find that the scalar-field approximation predicts the same leading suppression of small velocity differences near every simple nonzero Genesis solution. Including gravitational backreaction changes this finite-time suppression and gives different results for the two real Hubble solutions allowed by the background equation. We also show that homogeneous convergence may occur near both perturbatively healthy and unstable backgrounds, and that strong suppression in the velocity variable does not necessarily imply comparable suppression in the Hubble parameter. For the contracting model, we derive the sensitivity analytically and find that the physically relevant power-law solution attracts nearby homogeneous trajectories, whereas the second power-law solution repels them.

The paper is organized as follows. Section~\ref{sec:Horndeski_Genesis_Bounce} introduces the theoretical framework and the relevant cosmological backgrounds. Section~\ref{sec:The_General_Reduction_Scheme} develops the reduction procedure. Section~\ref{sec:Models} presents the three models and examines their phase portraits and perturbative properties. Section~\ref{sec:Initial_Conditions} quantifies the dependence on the initial velocity and the conditioning of its inverse reconstruction. Finally, Section~\ref{sec:Conclusions} summarizes the results and discusses their limitations.

%%%%%%%%%%%%%%%%%%%%%%%%%%%%%%%%%%%%%%%%%%%%%%%%%%
\section{The Galilean Genesis and the bouncing Universe within the Horndeski theory}
\label{sec:Horndeski_Genesis_Bounce}
%%%%%%%%%%%%%%%%%%%%%%%%%%%%%%%%%%%%%%%%%%%%%%%%%%

\subsection{Cosmological background equations}

We consider Horndeski theory, the most general four-dimensional scalar--tensor theory constructed from a metric $g_{\mu \nu}$ and a single scalar field $\phi$ whose Euler--Lagrange equations contain derivatives of at most second order (\cite{Horndeski:1974wa}, see also~\cite{Kobayashi:2019hrl} for a review). We adopt the metric signature $(-, +, +, +)$ and consider the scalar--gravity system without additional matter components. Its action can be written as
\begin{equation}
    %OK!
    \label{eq:Horndeski_Action}
    S=\int \dd[4]{x} \sqrt{-g} \sum\limits_{i = 2}^{5} \mathcal{L}_i,
\end{equation}
where
\begin{align}
    %OK! Compared to 1501.02553v2
    \mathcal{L}_2 &= G_2(\phi, X), \\
    \mathcal{L}_3 &= -G_3(\phi, X) \Box\phi, \\
    \mathcal{L}_4 &= G_4(\phi, X) R +G_{4 X} \left[(\Box \phi)^2 - (\nabla_\mu \nabla_\nu \phi)^2\right], \\
    \mathcal{L}_5 &= G_5(\phi, X) G^{\mu \nu} \nabla_\mu \nabla_\nu \phi - \frac{G_{5 X}}{6} \left[(\Box\phi)^3 - 3\Box\phi (\nabla_\mu \nabla_\nu \phi)^2 + 2(\nabla_\mu \nabla_\nu \phi)^3\right].
\end{align}
Here $g \coloneq \det g_{\mu \nu}$, $\nabla_\mu$ is the Levi--Civita covariant derivative associated with $g_{\mu \nu}$, $R$ is the Ricci scalar, $G_{\mu \nu}$ is the Einstein tensor, and $\Box \phi \coloneq g^{\mu \nu} \nabla_\mu \nabla_\nu \phi$. Our curvature convention is such that, on the FLRW background introduced below, $R = 6 (\dot{H} + 2 H^2)$. The scalar kinetic variable is defined by
\begin{equation}
    %OK!
    X \coloneq -\frac{1}{2} g^{\mu \nu} \partial_\mu \phi \partial_\nu \phi.
\end{equation}
The contractions appearing in the action are
\begin{align}
    %OK!
    (\nabla_\mu \nabla_\nu \phi)^2 &\coloneq (\nabla_\mu \nabla_\nu \phi) (\nabla^\mu \nabla^\nu \phi), \\
    (\nabla_\mu \nabla_\nu \phi)^3 &\coloneq (\nabla_\mu \nabla_\nu \phi) (\nabla^\nu \nabla^\rho \phi) (\nabla_\rho \nabla^\mu \phi).
\end{align}
The four arbitrary functions $G_i(\phi, X)$ are assumed to be sufficiently differentiable on the domain under consideration. Subscripts $\phi$ and $X$ denote partial derivatives with respect to the corresponding arguments; in particular, $G_{i X} \coloneq \pdv*{G_i}{X}$. The gravitational normalization is included in these functions: for $G_4 = M_{\mathrm{Pl}}^2/2$ and $G_5=0$, $\mathcal{L}_4$ reduces to the Einstein--Hilbert Lagrangian, where $M_{\mathrm{Pl}}$ is the reduced Planck mass.

We restrict our analysis to a spatially flat Friedmann--Lema\^itre--Robertson--Walker (FLRW) background,
\begin{equation}
    %OK!
    \label{eq:FLRW_Metric}
    \dd{s}^2 = -\dd{t}^2 + a^2(t) \delta_{ij} \dd{x^i} \dd{x^j},
\end{equation}
with a spatially homogeneous scalar field,
\begin{equation}
    %OK!
    \phi=\phi(t).
\end{equation}
Here $t$ denotes cosmic time, $a(t)$ is the scale factor, and $H \coloneq \dot{a}/a$ is the Hubble parameter. A dot denotes differentiation with respect to cosmic time $t$.

For this ansatz, the temporal and spatial components of the metric field equations reduce to the Friedmann constraint $\mathcal{E} = 0$ and the evolution equation $\mathcal{P} = 0$, respectively~\cite{Kobayashi:2011nu}:
\begin{equation}
    \label{eq:The_Friedmann_Equation}
    \begin{split}
        \mathcal{E} &\coloneq 2XG_{2X}- G_2 +6X\dot\phi HG_{3X}-2XG_{3\phi} -6 H^2 G_4 \\
        &+ 24 H^2 X (G_{4X} + XG_{4XX}) - 12 H X \dot{\phi} G_{4 \phi X} - 6 H \dot{\phi} G_{4 \phi} \\
        &+ 2 H^3 X \dot{\phi} (5G_{5X} + 2XG_{5XX}) - 6H^2X (3 G_{5\phi} + 2 X G_{5\phi X}),
    \end{split}
\end{equation}
\begin{equation}
    %OK! Compared to 1501.02553v2
    \label{eq:The_Evolution_Equation}
    \begin{split}
        \mathcal{P} &\coloneq G_2 - 2 X (G_{3 \phi} + \ddot{\phi} G_{3X}) + 2 (3H^2 + 2 \dot{H}) G_4 - 12H^2 X G_{4X} -4 H \dot{X} G_{4 X} \\
        &- 8 \dot{H} X G_{4 X} - 8 H X \dot{X} G_{4 X X} + 2 (\ddot{\phi} + 2 H \dot{\phi}) G_{4 \phi} + 4 X G_{4\phi \phi} \\
        &+ 4 X (\ddot{\phi} - 2 H \dot{\phi}) G_{4\phi X} - 2 X (2 H^3 \dot{\phi} + 2 H \dot{H} \dot{\phi} + 3 H^2 \ddot{\phi}) G_{5 X} - 4 H^2 X^2 \ddot{\phi} G_{5 X X} \\
        &+ 4 H X (\dot{X} - H X) G_{5\phi X} + 2 \left[ 2\dv{t} (H X) + 3 H^2 X \right] G_{5 \phi} + 4 H X \dot{\phi} G_{5 \phi \phi}.
    \end{split}
\end{equation}
All functions and their derivatives are evaluated at $\phi = \phi(t)$ and $X = \dot{\phi}^2/2$. The quantities $\mathcal{E}$ and $\mathcal{P}$ include the gravitational contributions and do not represent separately defined matter energy density and pressure.

Variation with respect to $\phi$ yields the scalar-field equation. Diffeomorphism invariance implies a background identity relating this equation to the two metric equations above. For the rolling-scalar backgrounds considered below, with $\dot{\phi} \neq 0$, the scalar equation follows from $\mathcal{E} = 0$ and $\mathcal{P} = 0$. We therefore take these two equations of motion~\cref{eq:The_Friedmann_Equation,eq:The_Evolution_Equation} as the independent equations governing the background evolution~\cite{Kobayashi:2019hrl}. At isolated scalar turning points with $\dot{\phi}=0$, this implication need not hold, because the background identity relating the scalar equation to the metric equations no longer permits one to recover the scalar equation by division by $\dot{\phi}$. Such points therefore require separate treatment in the reduced phase-space description developed below.

%%%%%%%%%%%%%%%%%%%%%%%%%%%%%%%%%%%%%%%%%%%%%%%%%%%%%%
\subsection{Galilean Genesis and bouncing cosmologies}
%%%%%%%%%%%%%%%%%%%%%%%%%%%%%%%%%%%%%%%%%%%%%%%%%%%%%%

We are interested in two classes of nonstandard cosmological histories: Galilean Genesis and nonsingular bouncing evolution. In the Galilean Genesis scenario, the Universe approaches Minkowski spacetime in the asymptotic past. A characteristic expanding Genesis background satisfies
\begin{equation}
    %Should be OK!
    \label{eq:Genesis_Asymptotics_General}
    \begin{split}
        a(t) &\to a_{-\infty} > 0, \\
        H(t) &\to 0^+, \\
        \dot H(t) &\to 0^+,
    \end{split}
\end{equation}
as $t \to -\infty$, while $\dot{H} > 0$ during the Genesis stage. Unlike a bouncing cosmology, Genesis does not require a preceding contracting phase. In generalized Galilean Genesis, the nontrivial derivative interactions of the scalar field allow this type of evolution to arise within the Horndeski theory, see~\cite{Nishi:2015pta}.

A bouncing cosmology, on the other hand, connects a contracting phase to an expanding one. At a regular, nondegenerate bounce at $t = t_b$, the background obeys
\begin{equation}
    %Should be OK!
    \label{eq:Bounce_Conditions}
    \begin{split}
        a(t_b) &> 0, \\
        H(t_b) &= 0, \\
        \dot{H}(t_b) &> 0.
    \end{split}
\end{equation}
In a neighborhood of $t_b$, the Hubble parameter passes from negative to positive values, and the scale factor has a nonzero local minimum. These local conditions do not, by themselves, establish the regularity of the entire cosmological history.

In the present work, however, we analyze only the contracting stage that precedes the bounce, rather than the transition through the bounce itself. The background considered below has the power-law form
\begin{equation}
    \label{eq:Contraction_Properties}
    \begin{split}
        a(t) &= d(-t)^\chi > 0, \\
        H(t) &= \frac{\chi}{t} < 0, \\
        \dot H(t) &= -\frac{\chi}{t^2} < 0,
    \end{split}
    \qquad
    t<0,\quad 0<\chi<1.
\end{equation}
Thus, during the interval in which this power-law description is valid, the scale factor decreases monotonically as cosmic time increases, while the magnitude of the negative Hubble parameter grows. This contracting solution describes the stage preceding the bounce and does not itself contain the transition to expansion. Related Horndeski models in which contraction is continued through a stable nonsingular bounce to later expanding stages, including inflationary and general-relativistic kination epochs, were constructed in~\cite{Ageeva:2024vtr,Ageeva:2022asq}.

Both scenarios of Genesis or Bounce thus involve a stage with $\dot{H} > 0$. In general relativity, for a spatially flat FLRW spacetime sourced by minimally coupled isotropic matter, the Einstein equations give
\begin{equation}
    %OK!
    \dot{H} = -\frac{\rho + p}{2 M_{\mathrm{Pl}}^2},
\end{equation}
where $\rho$ and $p$ denote the matter energy density and pressure. Therefore, in this setting, $\dot{H} > 0$ is equivalent to $\rho + p < 0$ and hence to violation of the null energy condition~\cite{Rubakov:2014jja}.

In Horndeski theory, the gravitational field equations do not generally take the Einstein form, and their separation into a conventional gravitational sector and an effective energy--momentum tensor assigned to the scalar field is not unique. Consequently, the condition $\dot{H} > 0$ cannot in general be identified unambiguously with violation of the null energy condition by an effective scalar-field component. It nevertheless provides a useful geometric criterion for identifying Genesis-type evolution and a necessary local condition for a regular, nondegenerate bounce.

The existence of a background solution must be distinguished from its perturbative stability. In particular, the absence of ghost and gradient instabilities must be verified independently in both the scalar and tensor sectors. Moreover, no-go theorems show that, under their respective assumptions, spatially flat cosmological solutions that remain nonsingular and stable throughout their entire evolution cannot generically be realized within Horndeski theory~\cite{Libanov:2016kfc,Kobayashi:2016xpl}. These results do not by themselves exclude a regular background solution over a finite interval, but they must be taken into account when assessing whether that solution admits a complete and physically viable cosmological history. In the present work, we first analyze the homogeneous background dynamics; perturbative stability must be examined separately.

%%%%%%%%%%%%%%%%%%%%%%%%%%%%%%%%%%%%%%
\section{The general reduction scheme}
\label{sec:The_General_Reduction_Scheme}
%%%%%%%%%%%%%%%%%%%%%%%%%%%%%%%%%%%%%%

In this section, we study a unified procedure for constructing phase portraits of homogeneous cosmological solutions in a subclass of Horndeski theory. The choice of phase-space variables and the general logic of the construction are motivated by the analysis of generalized Galilean Genesis in~\cite{Nishi:2015pta}, where homogeneous solutions were studied in the $(\psi, \dot{\psi})$ plane with $\psi= e^{-\lambda \phi}$. We adapt this phase-space description to the full background equations and formulate it in a common form applicable to the two Genesis models considered below and to the contracting stage of a bouncing cosmology. Although their Lagrangian functions differ, all three models satisfy 
\begin{equation}
    \label{eq:Reduction_Common_Subclass}
    G_4 = G_4(\phi), \qquad G_5=0.
\end{equation}
For $G_4 \neq 0$, these restrictions make the Friedmann constraint quadratic in the Hubble parameter $H$. We introduce suitable phase-space variables and use this constraint to construct a reduced dynamical system on each branch wherever the two roots are real and distinct. We note, that the points at which the roots coincide require separate consideration.

%%%%%%%%%%%%%%%%%%%%%%%%%%%%%%%%%%%%%%%%%%%%%%%%%%%%%%%%%%%%%%%
\subsection{Phase-space variables and the Friedmann constraint}
\label{subsec:Phase-Space_Variables}
%%%%%%%%%%%%%%%%%%%%%%%%%%%%%%%%%%%%%%%%%%%%%%%%%%%%%%%%%%%%%%%

We introduce the field redefinition, following~\cite{Nishi:2015pta}
\begin{equation}
    %OK!
    \label{eq:Reduction_Variables}
    \psi\coloneq e^{-\lambda\phi},\qquad v\coloneq\dot\psi,
\end{equation}
where $\lambda \neq 0$ and $\psi > 0$. In these variables, the scalar field and its time derivatives are
\begin{align}
    %OK! Consistent with the calculations
    \label{eq:Reduction_Phi_Relations}
    \phi &= -\frac{\ln \psi}{\lambda}, &\dot{\phi} &= -\frac{v}{\lambda \psi}, &\ddot{\phi} &= \frac{v^2}{\lambda \psi^2} - \frac{\dot{v}}{\lambda \psi},\\
    \label{eq:Reduction_X_Relations}
    X &= \frac{v^2}{2 \lambda^2 \psi^2}, &\dot{X} &= -\frac{v}{\lambda^2 \psi^2} \left(\frac{v^2}{\psi} - \dot{v}\right).
\end{align}
The parameter $\lambda$ specifies the choice of phase-space coordinates; its value is fixed separately for each example. In particular, the same transformation can be used even when the functions $G_i$ involve parameters other than $\lambda$ in their exponential dependence on $\phi$.

For any specified Horndeski functions, the Friedmann equation~\cref{eq:The_Friedmann_Equation} can now be expressed as the algebraic constraint
\begin{equation}
    \label{eq:Reduction_Constraint}
    \mathcal{E}(\psi, v, H) = 0.
\end{equation}
In general Horndeski theory, this constraint is at most cubic in $H$. Wherever the cubic coefficient is nonzero, at least one real root exists, so the sign of the discriminant does not determine whether real solutions exist. If the polynomial degree decreases, the resulting constraint must be examined separately. In the subclass~\cref{eq:Reduction_Common_Subclass}, the constraint is quadratic whenever $G_4 \neq 0$, and its discriminant determines the domain admitting real roots.

%%%%%%%%%%%%%%%%%%%%%%%%%%%%%%%%%%%%%%%%%%%%%%%%%%%%%%%%%%%%%%%%%%%%%%%%%%%%%%%%%%%%%%%%%%%%%%%%%%%%%%%%%%%%%%
\subsection{Quadratic Friedmann constraint for \texorpdfstring{$G_4 = G_4(\phi)$ and $G_5 = 0$}{G4(phi) and G5=0}}
%%%%%%%%%%%%%%%%%%%%%%%%%%%%%%%%%%%%%%%%%%%%%%%%%%%%%%%%%%%%%%%%%%%%%%%%%%%%%%%%%%%%%%%%%%%%%%%%%%%%%%%%%%%%%%

For the common subclass~\cref{eq:Reduction_Common_Subclass}, the independent background equations simplify to
\begin{align}
    %OK! These simplifications reverified by me
    \label{eq:Reduction_E_Subclass}
    &2 X G_{2X} - G_2 - 2 X G_{3\phi} +6 H \dot{\phi}\left(X G_{3X} - G_{4\phi}\right) - 6 H^2 G_4 = 0, \\
    \label{eq:Reduction_P_Subclass}
    &G_2 - 2 X \left(G_{3 \phi} + \ddot{\phi} G_{3X}\right) + 2 \left(3H^2 + 2 \dot{H}\right) G_4\nonumber\\
    &+ 2 \left(\ddot{\phi} + 2 H \dot{\phi}\right) G_{4 \phi} + 4 X G_{4\phi \phi} = 0.
\end{align}
Here and below, all functions and their derivatives are evaluated at $\phi = -\ln \psi/\lambda$ and $X = v^2/(2 \lambda^2 \psi^2)$ after the field redefinition, see~\cref{eq:Reduction_Variables}.
The equation~\cref{eq:Reduction_E_Subclass} takes the form
\begin{equation}
    %OK!
    \label{eq:Reduction_Quadratic}
    A(\psi, v) H^2 + B(\psi, v) H + C(\psi, v) = 0,
\end{equation}
with coefficients
\begin{align}
    \label{eq:Reduction_Coefficients}
    A &= - 6 G_4, \\
    B &= 6 \dot{\phi} \left(X G_{3 X} - G_{4 \phi}\right), \\
    C &= 2 X G_{2X} - G_2 - 2 X G_{3 \phi}.
\end{align}
Whenever $A \neq 0$, the two algebraic roots are
\begin{equation}
    \label{eq:Reduction_H_Branches}
    H_{\pm}(\psi, v) = \frac{-B \pm \sqrt{\mathcal{D}}}{2 A}, \qquad \mathcal{D}(\psi, v) \coloneq B^2 - 4 A C.
\end{equation}
Thus, $\mathcal{D} > 0$ corresponds to two distinct real branches, $\mathcal{D} = 0$ to a repeated root, and $\mathcal{D} < 0$ to the absence of a real solution of the Friedmann constraint. This last region is excluded from the physical background phase space. The set $\mathcal{D}=0$ requires separate analysis because the two algebraic roots coincide there, and the corresponding solutions cannot in general be continued independently using the explicit root formula. The labels $\pm$ refer to the sign preceding the square root and do not specify whether the corresponding solution is expanding or contracting; this is determined by the sign of $H_\pm$.

The two Genesis examples have $G_4 = M_{\mathrm{Pl}}^2/2$, whereas the contracting-stage model has a nonconstant $G_4(\phi)$. Both cases are described by~\cref{eq:Reduction_E_Subclass,eq:Reduction_Coefficients}; the latter would not be covered by a reduction restricted to constant $G_4$.

%%%%%%%%%%%%%%%%%%%%%%%%%%%%%%%%%%%%%%%%%%%%%%%%%%%%%%%%%%
\subsection{Evolution equations on the algebraic branches}
%%%%%%%%%%%%%%%%%%%%%%%%%%%%%%%%%%%%%%%%%%%%%%%%%%%%%%%%%%

After substituting~\cref{eq:Reduction_Phi_Relations,eq:Reduction_X_Relations}, the evolution equation~\cref{eq:Reduction_P_Subclass} is linear in $\dot{v}$ and $\dot{H}$. We write it as
\begin{equation}
    %Checked!
    \label{eq:Reduction_P_Linear}
    \mathcal{P} = P_0(\psi, v, H) + P_v(\psi, v, H) \dot{v} + P_H(\psi, v, H) \dot{H} = 0,
\end{equation}
where $P_v$ and $P_H$ denote the coefficients of the time derivatives, not partial derivatives of $\mathcal{P}$.
Introducing
\begin{equation}
    %OK!
    U \coloneq \frac{v^2}{\lambda \psi^2}, \qquad W \coloneq \frac{1}{\lambda \psi}, \qquad K \coloneq X G_{3 X} - G_{4 \phi},
\end{equation}
so that $\ddot{\phi} = U - W \dot{v}$, we find
\begin{align}
 \label{eq:Reduction_P0}
    P_0 &= G_2 - 2 X (G_{3 \phi} + U G_{3 X}) + 6H^2 G_4 + 2(U + 2 H \dot{\phi}) G_{4 \phi} + 4 X G_{4\phi \phi},\\
    \label{eq:Reduction_P_Coefficients}
    P_v &= 2 W K = \frac{2}{\lambda \psi} \left(X G_{3 X} - G_{4 \phi}\right), \\
    P_H &= 4 G_4.
\end{align}
In the constant-$G_4$ limit, these expressions reduce to $P_H = 2M_{\mathrm{Pl}}^2$ and $P_v = 2 X G_{3 X}/(\lambda \psi)$, as used for the Genesis examples.

Since the first Friedmann~\cref{eq:Reduction_E_Subclass} must hold at every time, its derivative along a homogeneous trajectory satisfies
\begin{equation}
    \label{eq:Reduction_dE}
    \dv{\mathcal{E}}{t} = \mathcal{E}_\psi v + \mathcal{E}_v \dot{v} + \mathcal{E}_H \dot{H} = 0,
\end{equation}
where $\mathcal{E}_\psi$, $\mathcal{E}_v$ and $\mathcal{E}_H$ denote partial derivatives of $\mathcal{E}(\psi, v, H)$, with the remaining variables kept fixed. Together with~\cref{eq:Reduction_P_Linear}, this gives
\begin{equation}
    \label{eq:Reduction_Linear_System}
    \begin{pmatrix}
        \mathcal{E}_v & \mathcal{E}_H \\
        P_v & P_H
      \end{pmatrix}
    \begin{pmatrix}
        \dot{v} \\
        \dot{H}
    \end{pmatrix}
    =
    -
    \begin{pmatrix}
        \mathcal E_\psi v \\
        P_0
    \end{pmatrix}.
\end{equation}
The corresponding determinant is
\begin{equation}
    \label{eq:Reduction_Evolution_Determinant}
    \Delta \coloneq \mathcal{E}_v P_H - \mathcal{E}_H P_v.
\end{equation}

For the subclass $G_4=G_4(\phi)$ and $G_5=0$, the determinant
contains an explicit factor of $v$. Indeed, using
$K=XG_{3X}-G_{4\phi}$, one finds
\begin{equation}
    \label{eq:Reduction_Delta_Factorization}
    \Delta
    =
    \frac{4v}{\lambda^2\psi^2}\,
    \mathcal{K}_{\mathrm{red}},
\end{equation}
where
\begin{equation}
    \label{eq:Reduction_Kred}
    \begin{split}
    \mathcal{K}_{\mathrm{red}}
    \coloneq{}&
    G_4\left[
        G_{2X}
        +2XG_{2XX}
        -2G_{3\phi}
        -2XG_{3\phi X}
    \right.
    \\
    &\left.
        \hspace{18mm}
        +6H\dot{\phi}
        \left(G_{3X}+XG_{3XX}\right)
    \right]
    +3\left(XG_{3X}-G_{4\phi}\right)^2 .
    \end{split}
\end{equation}
Consequently, if the coefficient functions and their derivatives remain finite at $X = 0$, then
\begin{equation}
    v=0 \quad\Longrightarrow\quad \Delta=0.
\end{equation}
This automatic zero does not by itself signal a singularity of the original background equations. Since
\begin{equation}
    \dot{\phi} = -\frac{v}{\lambda\psi},
\end{equation}
the locus $v = 0$ is a scalar turning point, $\dot{\phi} = 0$. The background diffeomorphism identity can be written as
\begin{equation}
    \dot{\mathcal{E}} + 3 H \left(\mathcal{E} + \mathcal{P} \right) = \dot{\phi} \mathcal{C}_\phi,
\end{equation}
where $\mathcal{C}_\phi=0$ denotes the scalar-field equation. For $\dot{\phi}\neq0$, the equations $\mathcal{E} = \mathcal{P} =0$ imply $\mathcal{C}_\phi=0$. At $\dot{\phi}=0$, however, the same identity reduces to $0 = 0$ and supplies no scalar equation. Therefore, replacing the scalar equation by the differentiated constraint~\cref{eq:Reduction_dE} loses independent scalar information at $v = 0$. Continuation through such a point must be examined using the original scalar equation together with an independent metric equation, rather than the formulas below.

For $\Delta \neq 0$, the derivatives are determined uniquely:
\begin{align}
    %OK!
    \label{eq:Reduction_vdot}
    \dot{v} &= \frac{\mathcal{E}_H P_0 - \mathcal{E}_\psi v P_H}{\Delta}, \\    \label{eq:Reduction_Hdot}
    \dot{H} &= \frac{\mathcal{E}_\psi v P_v - \mathcal{E}_v P_0}{\Delta}.
\end{align}
Define
\begin{equation}
    %OK!
    F(\psi, v, H) \coloneq \frac{\mathcal{E}_H P_0 - \mathcal{E}_\psi v P_H}{\Delta}.
\end{equation}
Substitution of the Friedmann roots yields the branch-dependent autonomous systems
\begin{equation}
    %OK!
    \label{eq:reduction_vector_fields}
    \dot{\psi} = v, \quad \dot{v} = F_\pm(\psi, v), \quad F_\pm(\psi, v) \coloneq F(\psi, v, H_\pm(\psi, v)).
\end{equation}
These branchwise systems are defined on domains where
$G_4\neq0$, $\mathcal D>0$, and $\Delta\neq0$, provided that the
coefficient functions are sufficiently smooth. The exceptional sets
$\mathcal D=0$ and $\Delta=0$ have different mathematical and
physical meanings.

First, on either algebraic root,
\begin{equation}
    \label{eq:Reduction_EH_Discriminant}
    \mathcal E_H(\psi,v,H_\pm)
    =
    2A H_\pm+B
    =
    \pm\sqrt{\mathcal D}.
\end{equation}
Hence $\mathcal D=0$ is the locus at which the two algebraic roots
of the Friedmann constraint coalesce and $\mathcal E_H=0$. It is
generically a fold of the projection of the constraint surface
$\mathcal E(\psi,v,H)=0$ onto the $(\psi,v)$ plane: the constraint
can no longer be represented locally by two distinct graphs
$H_\pm(\psi,v)$. The unreduced background evolution may nevertheless
remain regular there, and a solution may pass through the fold while
the algebraic labels of the two roots are exchanged.

By contrast, $\Delta=0$ means that the coefficient matrix in
\eqref{eq:Reduction_Linear_System} is singular. The differentiated
constraint and the evolution equation then fail to determine
$(\dot v,\dot H)$ uniquely within this reduced formulation. At
$v=0$, this is the scalar-turning-point degeneracy discussed above.
Away from a scalar turning point, namely for $v\neq0$, and away from
a Friedmann fold, $\mathcal D>0$, the condition $\Delta=0$ is not a
coalescence of the algebraic Hubble roots.

More specifically, on the Friedmann constraint and for the present
Horndeski subclass, the scalar kinetic coefficient $\mathcal G_S$ in
the quadratic action satisfies
\begin{equation}
    \label{eq:Reduction_Delta_GS_Relation}
    \mathcal G_S\,\mathcal D
    =
    18G_4 v\,\Delta.
\end{equation}
Therefore, on a rolling regular branch with
$v\neq0$, $\mathcal D>0$, and $G_4\neq0$,
\begin{equation}
    \Delta=0
    \quad\Longleftrightarrow\quad
    \mathcal G_S=0.
\end{equation}
In that domain, $\Delta=0$ represents a kinetic degeneracy of the
scalar sector rather than a fold of the Friedmann constraint.
Neither $\mathcal D=0$ nor $\Delta=0$ implies, by itself, a spacetime
curvature singularity, but the two loci must be analyzed separately
using the unreduced background equations. The solutions of these systems define the projected cosmological trajectories in the $(\psi, v)$ plane. The corresponding Hubble parameter is recovered from~\cref{eq:Reduction_H_Branches}, and the scale factor can then be reconstructed from $\dot{a} = a H_\pm(\psi, v)$.

%%%%%%%%%%%%%%%%
\section{Models}
\label{sec:Models}
%%%%%%%%%%%%%%%%

In this section, we apply the general reduction scheme developed above to three different Horndeski models. Models~1 and~2 are generalized Galilean Genesis ``benchmark'' solutions introduced in~\cite{Nishi:2015pta}. We reproduce their phase portraits to illustrate and validate the reduction procedure, and then examine the corresponding algebraic branches, admissible phase-space domains, distinguished Genesis solutions, and their perturbative properties in greater detail. In Model~3, we apply the same scheme to the power-law contracting stage of the Horndeski bouncing construction of Ref.~\cite{Ageeva:2022asq}. This third example allows us to study the homogeneous phase-space structure preceding the bounce. The functions considered here describe only the contracting stage and do not implement the subsequent transition to expansion.

%%%%%%%%%%%%%%%%%%%%%%%%%%%%%%%%%%%%%%%%%%%%%%%%%%
\subsection{Model 1: Generalized Galilean Genesis}
\label{sec:model1}
%%%%%%%%%%%%%%%%%%%%%%%%%%%%%%%%%%%%%%%%%%%%%%%%%%

Let us consider the generalized Galilean Genesis model introduced in~\cite{Nishi:2015pta}. The corresponding Horndeski functions are
\begin{align}
    %OK!
    \label{eq:G_2_function_from_1501.02553}
    G_2 &= e^{2(\alpha+1)\lambda\phi} g_2(Y), \\
    \label{eq:G_3_function_from_1501.02553}
    G_3 &= e^{2\alpha\lambda\phi} g_3(Y), \\
    \label{eq:G_4_function_from_1501.02553}
    G_4 &= \frac{M_{\mathrm{Pl}}^2}{2} +e^{2\alpha\lambda\phi} g_4(Y), \\
    \label{eq:G_5_function_from_1501.02553}
    G_5 &= e^{-2\lambda\phi} g_5(Y),
\end{align}
where
\begin{equation}
    %OK!
    Y\coloneq e^{-2\lambda \phi} X,\label{Y_X}
\end{equation}
and $\alpha (>0)$ is a dimensionless parameter. The parameter $\alpha$ controls the time dependence of the Hubble parameter during the Genesis phase. The generalized Genesis solution takes the asymptotic form
\begin{align}
    %OK! It is in the cited paper
    e^{\lambda\phi} &\simeq \frac{1}{\lambda \sqrt{2Y_0}} \frac{1}{(-t)}, \\
    H &\simeq \frac{h_0}{(-t)^{2\alpha+1}}, \quad  t \to -\infty,
\end{align}
where $Y \simeq Y_0 > 0$ and $h_0 > 0$. The corresponding scale factor behaves as
\begin{equation}
    %OK! It is too.
    a \simeq 1 + \frac{1}{2\alpha} \frac{h_0}{(-t)^{2\alpha}},
\end{equation}
so that the spacetime approaches Minkowski space in the asymptotic past. Here we take $\lambda > 0$ and normalize the asymptotic scale factor to unity. These expressions describe the early-time regime $t \to -\infty$, in which $H \abs{t} \ll 1$, rather than the entire interval $t < 0$.

For the model functions given by~\cref{eq:G_2_function_from_1501.02553,eq:G_3_function_from_1501.02553,eq:G_4_function_from_1501.02553,eq:G_5_function_from_1501.02553}, the asymptotic background equations are conveniently expressed in terms of \cite{Nishi:2015pta}
\begin{align}
    %OK! It is too.
    \mathcal{E} &\simeq e^{2 (\alpha + 1) \lambda \phi} \hat{\rho}(Y_0) \simeq 0, \\
    \mathcal{P} &\simeq 2 \mathcal{G}(Y_0) \dot{H} + e^{2 (\alpha + 1) \lambda \phi} \hat{p}(Y_0) \simeq 0
\end{align}
with
\begin{align}
    %OK! It is too.
    \hat{\rho}(Y) &\coloneq 2 Y g_2'(Y) - g_2(Y) -4 \lambda Y \left[\alpha g_3(Y) - Y g_3'(Y)\right], \\
    \hat{p}(Y) &\coloneq g_2(Y) - 4 \alpha \lambda Y g_3(Y) + 8 (2 \alpha + 1) \lambda^2 Y \left[\alpha g_4(Y) - Y g_4'(Y) \right], \\
    \mathcal{G}(Y) &\coloneq M_{\mathrm{Pl}}^2 - 4 \lambda Y (g_5(Y) + Y g'_5(Y)).
\end{align}
The constant $Y_0$ is determined by
\begin{equation}
    %OK!
    \hat{\rho}(Y_0) = 0.
\end{equation}
Provided that $\mathcal{G}(Y_0) > 0$, the leading-order evolution equation implies that an expanding Genesis solution with $\dot{H} > 0$ requires
\begin{equation}
    %OK!
    \hat{p}(Y_0) < 0.
\end{equation}
More generally, the sign of $\dot{H}$ is determined by the ratio $-\hat{p}(Y_0)/\mathcal{G}(Y_0)$.

We investigate the particular choice of the functions $g_i(Y)$ underlying the phase portrait shown in Fig.~2 of~\cite{Nishi:2015pta},
\begin{align}
    %OK!
    \label{eq:g_2_Function}
    g_2(Y) &= -Y+Y^2, \\
    \label{eq:g_3_Function}
    g_3(Y) &= Y, \\
    \label{eq:g_4_Function}
    g_4(Y) &= 0, \\
    \label{eq:g_5_Function}
    g_5(Y) &= 0.
\end{align}
and thus one has
\begin{align}
    %OK!
    G_4 &= \frac{M_{\mathrm{Pl}}^2}{2}, \\
    G_5 &= 0,
\end{align}
and therefore this model belongs to the subclass discussed in the previous section. In particular, the Friedmann equation is quadratic in $H$ and gives two branches, $H_\pm$.

For the choice~\cref{eq:g_2_Function,eq:g_3_Function,eq:g_4_Function,eq:g_5_Function}, one obtains
\begin{equation}
    %OK! Checked by WM
    \hat{\rho}(Y) = - Y + \left[3 - 4 \lambda (\alpha-1) \right] Y^2.
\end{equation}

Apart from the trivial Minkowski root $Y = 0$, the equation $\hat{\rho}(Y_0) = 0$ yields
\begin{equation}
    %OK! Checked by me
    Y_0 = \frac{1}{3 - 4 \lambda (\alpha - 1)}.
\end{equation}
Consequently, the existence of a nontrivial Genesis background with $Y_0 > 0$ requires
\begin{equation}
    %OK! Checked by me
    3 - 4 \lambda (\alpha - 1) > 0.
\end{equation}
For $0 < \alpha \leq 1$, this condition is satisfied for any $\lambda > 0$, while for $\alpha > 1$ it gives
\begin{equation}
    %OK! Checked by me
    0 < \lambda < \frac{3}{4(\alpha-1)}.
\end{equation}
For the same model,
\begin{equation}
    %OK! Checked by me
    \hat{p}(Y_0) = -\frac{2 (1 + 2 \lambda)}{\left[3 - 4 \lambda (\alpha - 1)\right]^2},
\end{equation}
and hence $\hat{p}(Y_0) < 0$ for $\lambda > 0$. Thus, no additional restriction on the existence of the expanding Genesis background follows from this condition.

Using the variables~\cref{eq:Reduction_X_Relations,Y_X} introduced above, one may write
\begin{equation}
    %OK!
    Y = \frac{v^2}{2 \lambda^2},
\end{equation}
and so the asymptotic Genesis trajectory corresponds to
\begin{equation}
    %OK!
    v_{\mathrm{G}} = -\lambda \sqrt{\frac{2}{3 -4 \lambda (\alpha - 1) }}.
\end{equation}

The negative sign selects the sector $v < 0$, in which $\phi$ increases with cosmic time. The horizontal line $v = v_{\mathrm{G}}$ specifies the leading asymptotic Genesis velocity. In the full system including gravitational backreaction, the Genesis trajectory generally contains subleading corrections and is not an exactly invariant horizontal line. The homogeneous attraction toward Genesis-type evolution was established in~\cite{Nishi:2015pta}.

Since the Friedmann equation is quadratic in $H$, a real cosmological solution exists only if its discriminant is non-negative. In the present model the discriminant is
\begin{equation}
    %OK! True
    \mathcal{D} = \frac{3v^2}{\lambda^6 \psi^{2 + 4\alpha}} \left[3 v^4 + M_{\mathrm{Pl}}^2 \lambda^2 \psi^{2 \alpha} \left( \left[3 - 4 \lambda (\alpha - 1) \right]v^2 - 2 \lambda^2 \right) \right].
\end{equation}
The region with $\mathcal{D} < 0$ is therefore excluded from the physical phase space.

Evaluating the discriminant on the line $v = v_{\mathrm G}$ gives
\begin{equation}
    %OK!
    \eval{\mathcal{D}}_{\mathrm{G}} = \frac{72}{\left[3 - 4 \lambda (\alpha - 1) \right]^3 \psi^{2 + 4 \alpha}}.
\end{equation}
Thus, this line lies in the domain admitting two distinct real Friedmann roots for every finite $\psi > 0$. The nonzero-velocity boundary of the forbidden region approaches $v = v_{\mathrm{G}}$ as $\psi \to \infty$. This evaluation concerns the leading asymptotic line; the discriminant along the full Genesis trajectory must include the subleading correction to $v$.

For $3 - 4 \lambda (\alpha - 1) > 0$, the evolution determinant $\Delta$ has the sign of $v$ on both real Friedmann branches whenever $v \neq 0$. Thus, $\Delta$ imposes no additional restriction on the rolling region with $\mathcal{D} > 0$ in this model.

In the linearly stable parameter region identified below, where $4 \alpha \lambda>1$, the expanding Genesis asymptotics is associated with the $H_+$ branch under our branch convention. This identification requires the subleading gravitational correction to $v$: substituting $v = v_{\mathrm G}$ exactly into the Friedmann constraint at finite $\psi$ is insufficient to determine the branch. The $H_-$ solution is displayed for comparison and should not be interpreted as a second realization of the same expanding Genesis asymptotics.

\begin{figure}
    \centering
    \includegraphics[width=0.8\linewidth]{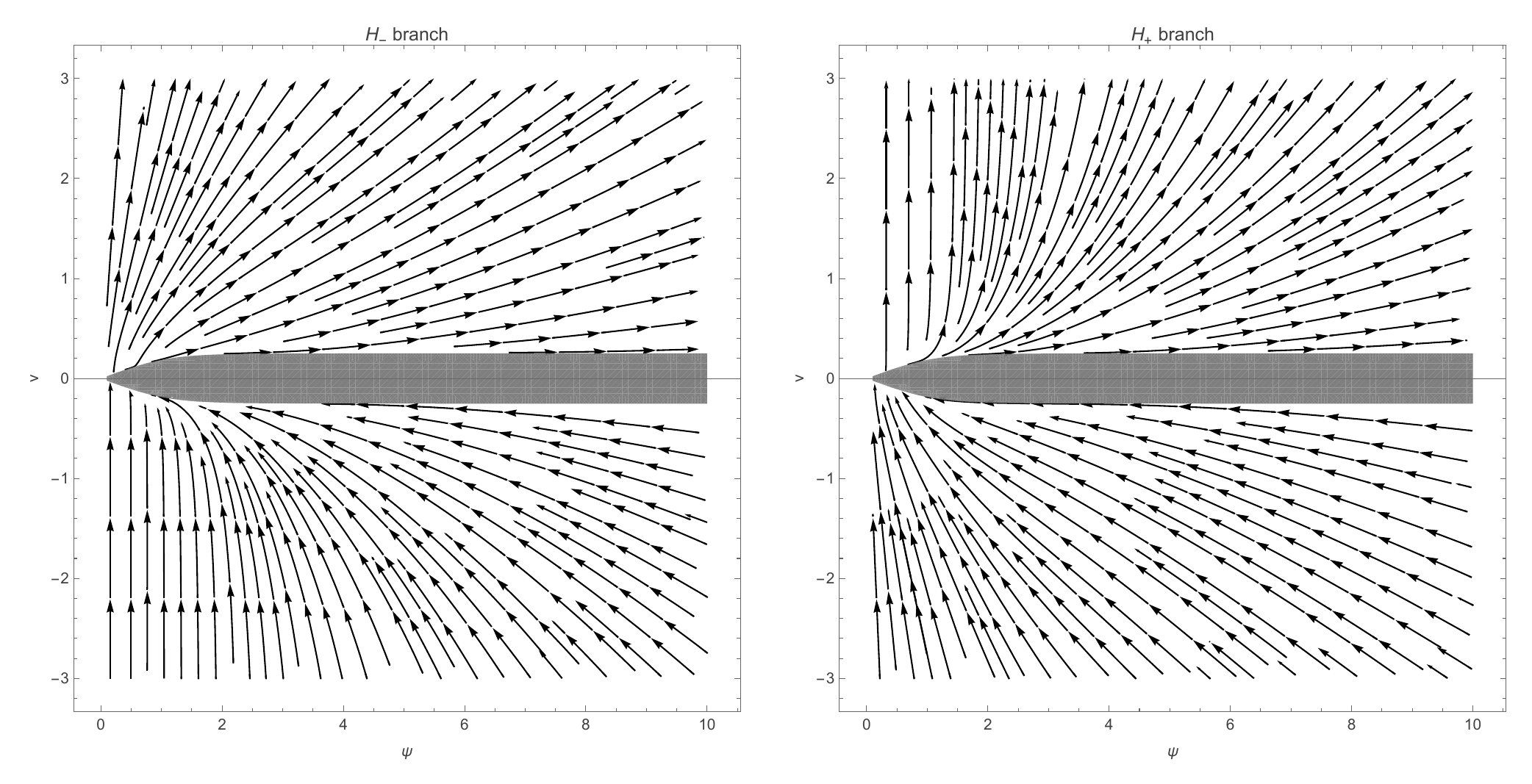}
    \caption{Phase portraits for the model of Genesis with~\cref{eq:g_2_Function,eq:g_3_Function,eq:g_4_Function,eq:g_5_Function} and for $\alpha = 2.02$, $\lambda = 0.25$, and $M_{\mathrm{Pl}} = 1$: the $H_-$ branch (left) and the $H_+$ branch (right). Grey shading marks the forbidden region $\mathcal{D} < 0$. Arrows indicate increasing cosmic time; in the sector $v < 0$, evolution proceeds toward decreasing $\psi$. The leading asymptotic Genesis velocity is $v_{\mathrm G} \simeq -0.251$.}
    \label{fig:1}
\end{figure}

As an illustration, we construct the phase portrait for $\alpha = 2.02$, $\lambda = 0.25$, and $M_{\mathrm{Pl}} = 1$, shown in~\cref{fig:1}. For these parameters, the asymptotic Genesis solution corresponds to $v_{\mathrm G} \simeq -0.251$. The chosen parameter values will also be shown below to satisfy the linear stability conditions and to yield a nearly scale-invariant scalar spectrum.

The existence of the Genesis background does not by itself guarantee its perturbative stability. Following the analysis of Ref.~\cite{Nishi:2015pta}, stability with respect to scalar perturbations requires
\begin{equation}
    %OK! From cited work
    \hat{\rho}'(Y_0) > 0,
\end{equation}
and
\begin{equation}
    %OK! From cited work
    \xi'(Y_0) < 0,
\end{equation}
where
\begin{equation}
    %OK! From cited work
    \xi(Y) \coloneq -\frac{Y \mathcal{G}(Y)}{\hat{p}(Y)}.
\end{equation}
For the present model, $g_5 = 0$, and hence
\begin{equation}
    %OK!
    \mathcal{G}(Y) = M_{\mathrm{Pl}}^2.
\end{equation}
Using the expression for $\hat{\rho}(Y)$ obtained above, we find
\begin{equation}
    %OK!
    \hat{\rho}'(Y_0) = 1 > 0,
\end{equation}
so that the first scalar stability condition is automatically satisfied.

For the same model,
\begin{equation}
    %OK!
    \hat{p}(Y) = -Y + \left(1 - 4 \alpha \lambda \right) Y^2,
\end{equation}
and therefore
\begin{equation}
    %OK!
    \xi(Y) = \frac{M_{\mathrm{Pl}}^2}{1 - \left(1 - 4 \alpha \lambda \right) Y}.
\end{equation}
Its derivative is
\begin{equation}
    %OK!
    \xi'(Y) = \frac{M_{\mathrm{Pl}}^2 \left(1 - 4 \alpha \lambda \right)}{\left[1 - \left(1 - 4 \alpha \lambda \right) Y \right]^2}.
\end{equation}
Hence, the second stability condition $\xi'(Y_0) < 0$ reduces to
\begin{equation}
    %OK!
    4 \alpha \lambda > 1.
\end{equation}

The tensor stability conditions of~\cite{Nishi:2015pta} are automatically satisfied in the present model for $M_{\mathrm{Pl}}^2 > 0$, since $g_5 = 0$ and therefore $\mathcal{G}(Y_0) = M_{\mathrm{Pl}}^2$.

Thus, restricting to $\alpha > 0$ and $\lambda > 0$, the stable Genesis region is determined by
\begin{equation}
    %OK!
    \begin{cases}
    3 - 4 \lambda (\alpha - 1) > 0, \\
    4 \alpha \lambda > 1.
    \end{cases}
\end{equation}

For $0 < \alpha \leq 1$, this gives
\begin{equation}
    %OK!
    \lambda>\frac{1}{4\alpha},
\end{equation}
whereas for $\alpha > 1$ one obtains
\begin{equation}
    %OK!
    \frac{1}{4\alpha} < \lambda < \frac{3}{4 (\alpha - 1)}.
\end{equation}

For the present model, the squared propagation speed of scalar perturbations takes the explicit form
\begin{equation}
    %OK!
    c_s^2 = \frac{4\alpha\lambda-1}{3-4\lambda(\alpha-1)}.
\end{equation}
Thus, the conditions derived above guarantee the absence of scalar ghost and gradient instabilities. If one additionally imposes subluminal scalar propagation, $c_s^2 \leq 1$, then, for $\alpha > 1/2$, one obtains
\begin{equation}
    %OK!
    \lambda \leq \frac{1}{2\alpha-1}.
\end{equation}

Having identified the parameter region in which the Genesis background is free from linear ghost and gradient instabilities, we now turn to its observational implications. Perturbative stability is necessary but not sufficient for a viable cosmological scenario: the primordial scalar and tensor perturbations must also be consistent with observational constraints~\cite{Planck:2019nip}. In the present subsection, we focus on the primordial power spectra. The asymptotic Genesis regime determines the spectral tilt, while the final amplitudes depend on the subsequent evolution and on the transition out of the Genesis phase.

According to~\cite{Nishi:2015pta}, the scalar spectral index in the generalized Genesis scenario is
\begin{equation}
    %OK! From cited paper.
    n_s =
    \begin{cases}
        3+2\alpha, & 0<\alpha<1/2,\\
        5-2\alpha, & \alpha>1/2.
    \end{cases}
\end{equation}
For $\alpha > 1/2$, the curvature perturbation possesses a growing superhorizon mode, $\zeta_k \propto(-t)^{1 - 2 \alpha}$. The corresponding spectral index is inherited by the final constant
curvature mode provided that the growing Genesis mode is transferred
with a nonzero, approximately $k$-independent coefficient over the
relevant long-wavelength range and dominates the final perturbation.

A nearly scale-invariant, slightly red-tilted scalar spectrum is therefore obtained for $\alpha$ slightly larger than $2$. For instance, $\alpha = 2.02$ gives $n_s = 0.96$. In this case, the stability conditions imply
\begin{equation}
    0.1238 < \lambda < 0.7353,
\end{equation}
while the additional requirement of subluminal scalar propagation restricts the interval to
\begin{equation}
    0.1238 < \lambda \leq 0.3289.
\end{equation}

The parameter choice used in~\cref{fig:1}, namely $\alpha = 2.02$ and $\lambda = 0.25$, lies within this interval. For these values, the squared scalar sound speed is $c_s^2 \simeq 0.515$, while the scalar spectral index is $n_s = 0.96$. This subluminality statement refers to the selected asymptotic Genesis
background and does not establish subluminality throughout the full
cosmological phase space. The distinction between background-local and
phase-space-wide subluminality, as well as the effect of an additional
matter sector, was analyzed for the model of
Ref.~\cite{Creminelli:2012my} in
Ref.~\cite{Easson:2013WhenMatter}.

Thus, the phase portrait illustrates the background dynamics for a parameter choice that satisfies the linear stability conditions in the asymptotic Genesis regime and yields a nearly scale-invariant scalar spectrum.

In the tensor sector, since $g_5 = 0$, the quadratic action reduces to
\begin{equation}
    %OK!
    S_h^{(2)} = \frac{M_{\mathrm{Pl}}^2}{8} \int \dd{t} \dd[3]{x} a^3 \left[\dot{h}_{ij}^2 - \frac{1}{a^2} (\partial_k h_{ij})^2\right].
\end{equation}
Consequently, tensor perturbations propagate with the speed of light, $c_T^2 = 1$, and are free from ghost and gradient instabilities. Since $a \simeq 1$ in the asymptotic Genesis regime, tensor vacuum fluctuations are not amplified, as shown in~\cite{Nishi:2015pta}. This absence of amplification applies to the leading asymptotic Genesis regime; tensor production during the subsequent transition requires a separate analysis.

Thus, the model admits linearly stable Genesis backgrounds with a
nearly scale-invariant scalar spectrum for an appropriate choice of
$\alpha$ and $\lambda$. However, the spectral tilt alone does not
establish full observational viability. The final perturbation
amplitudes depend on the completion of the Genesis phase. Moreover,
in the subclass with constant scalar and tensor propagation speeds
analyzed in Ref.~\cite{Akama:2022usl}, models that avoid strong
coupling and produce a slightly red-tilted scalar spectrum overproduce
scalar non-Gaussianity. For the purposes of this paper, the parameter
choice considered above therefore serves as a ``benchmark'' for
classical homogeneous dynamics and quadratic perturbative stability.

%%%%%%%%%%%%%%%%%%%%%%%%%%%%%%%%%%%%%%%%%%%%%%%%%%
\subsection{Model 2: Generalized Galilean Genesis with two Genesis solutions}
\label{sec:model2}
%%%%%%%%%%%%%%%%%%%%%%%%%%%%%%%%%%%%%%%%%%%%%%%%%%

We now consider the particular choice of the functions $g_i(Y)$ underlying the phase portrait shown in Fig.~3 of~\cite{Nishi:2015pta},
\begin{align}
    %OK!
    \label{eq:model2_g_2_Function}
    g_2(Y) &= -Y + 3Y^2 - Y^3, \\
    \label{eq:model2_g_3_Function}
    g_3(Y) &= Y, \\
    \label{eq:model2_g_4_Function}
    g_4(Y) &= 0, \\
    \label{eq:model2_g_5_Function}
    g_5(Y) &= 0.
\end{align}
This model differs from the model in~\cref{sec:model1} above only by the choice of $g_2(Y)$. Since $g_4 = g_5 = 0$, it belongs to the same subclass with $G_4 = M_{\mathrm{Pl}}^2/2$ and $G_5=0$. Consequently, the Friedmann equation remains quadratic in $H$ and admits two algebraic branches $H_\pm$.

For the choice~\cref{eq:model2_g_2_Function,eq:model2_g_3_Function,eq:model2_g_4_Function,eq:model2_g_5_Function}, one obtains
\begin{equation}
    \hat{\rho}(Y) = -Y + \left[ 9 - 4\lambda (\alpha - 1) \right] Y^2 - 5Y^3.
    %OK! Checked by WM
\end{equation}
It is convenient to introduce
\begin{equation}
    %OK!
    A_2 \coloneq 9 - 4 \lambda (\alpha - 1).
\end{equation}
Apart from the trivial Minkowski root $Y = 0$, the equation
\begin{equation}
    %OK!
    \hat{\rho}(Y_0) = 0
\end{equation}
has two solutions,
\begin{equation}
    Y_{0,\pm} = \frac{A_2 \pm \sqrt{A_2^2 - 20}}{10}.
    %OK! Checked by WM
\end{equation}
Thus, two distinct positive roots exist provided
\begin{equation}
    A_2 > 2 \sqrt{5}.
\end{equation}
At $A_2 = 2 \sqrt{5}$ the two roots coincide.

For the present model,
\begin{equation}
    \hat{p}(Y) = -Y + \left( 3 - 4 \alpha \lambda \right) Y^2 - Y^3.
    %OK! Checked by WM
\end{equation}
Using the equation satisfied by a nontrivial root $Y_0$,
\begin{equation}
    %OK!
    5 Y_0^2 - A_2 Y_0 + 1 = 0,
\end{equation}
this can be written as
\begin{equation}
    %OK!
    \hat{p}(Y_0) = -2 Y_0^2 (3 + 2 \lambda - 2 Y_0).
\end{equation}
The smaller root $Y_{0, -}$ always satisfies $\hat{p}(Y_{0, -}) < 0$ for $\lambda > 0$. The larger root $Y_{0, +}$ also describes an expanding Genesis background provided
\begin{equation}
    %OK!
    4 (4 \alpha + 1) \lambda^2 + 24 \alpha \lambda - 5 > 0.
\end{equation}

Using $Y = v^2/(2 \lambda^2)$, the two asymptotic Genesis solutions correspond to
\begin{equation}
    %OK!
    v_{\mathrm{G}, \pm} = -\lambda \sqrt{2 Y_{0,\pm}} = -\lambda \sqrt{\frac{A_2 \pm \sqrt{A_2^2 - 20}}{5}}.
\end{equation}
The negative sign selects the sector $v < 0$, in which $\phi$ increases with cosmic time. The horizontal lines $v = v_{\mathrm{G}, \pm}$ specify the leading asymptotic velocities of the two Genesis solutions. In the complete system, gravitational backreaction generally produces subleading corrections to $v$, so these lines need not be exactly invariant trajectories at finite $\psi$.

The discussion of the physical region proceeds exactly as in model of.~\cref{sec:model1}. The discriminant of the quadratic Friedmann equation is now
\begin{equation}
\label{D_mod2}
    %OK!
    \mathcal{D} = \frac{3v^2}{2 \lambda^6 \psi^{2 + 4 \alpha}} \left[6v^4 + M_{\mathrm{Pl}}^2 \psi^{2 \alpha} \left(2 A_2 \lambda^2 v^2 - 5 v^4 - 4 \lambda^4\right)\right],
\end{equation}
and the region with $\mathcal{D} < 0$ is excluded from the physical phase space.

On either asymptotic Genesis line, using $v_{\mathrm{G}, \pm}^2 = 2 \lambda^2 Y_{0, \pm}$ and $\hat{\rho}(Y_{0, \pm}) = 0$, the discriminant reduces to
\begin{equation}
    %OK!
    \eval{\mathcal{D}}_{\mathrm{G}, \pm} = \frac{72 Y_{0, \pm}^3}{\psi^{2 + 4 \alpha}} > 0
\end{equation}
for $\psi>0$.
Furthermore, the expression inside the square brackets in $\mathcal{D}$, see~\cref{D_mod2}, can be factorized as
\begin{equation}
    %OK!
    6 v^4 -5 M_{\mathrm{Pl}}^2 \psi^{2 \alpha}
    (v^2 - v_{\mathrm{G}, -}^2) (v^2 -v_{\mathrm{G}, +}^2).
\end{equation}
Consequently, as $\psi \to \infty$, the region admitting real Friedmann roots in the sector $v < 0$ approaches the interval
\begin{equation}
    %OK!
    v_{\mathrm{G}, +} \leq v \leq v_{\mathrm{G}, -}.
\end{equation}
The labels $\pm$ in $Y_{0,\pm}$ distinguish the two nontrivial roots of $\hat{\rho}(Y) = 0$ and should not be confused with the algebraic Friedmann branches $H_\pm$. For the benchmark parameters considered below and with the branch convention adopted above, both expanding Genesis asymptotics are continuously associated with the $H_+$ branch when the corresponding subleading gravitational corrections to $v$ are included. The $H_-$ branch is the companion algebraic solution and is displayed for comparison.

The perturbative stability conditions distinguish the two Genesis solutions. Evaluating the derivative of $\hat{\rho}(Y)$ at the two roots gives
\begin{equation}
    %OK!
    \hat{\rho}'(Y_{0, \pm}) = \mp Y_{0, \pm} \sqrt{A_2^2 - 20}.
\end{equation}
Thus, $\hat{\rho}'(Y_{0,-}) > 0$, whereas $\hat{\rho}'(Y_{0, +}) < 0$. The condition for the absence of scalar ghosts therefore selects the smaller root $Y_{0, -}$, while the solution associated with $Y_{0, +}$ is ghost-unstable.

Since $g_5 = 0$, one has $\mathcal{G}(Y) = M_{\mathrm{Pl}}^2$, and therefore
\begin{equation}
    %OK!
    \xi(Y) = \frac{M_{\mathrm{Pl}}^2}{1 - (3 - 4 \alpha \lambda) Y + Y^2}.
\end{equation}
Its derivative is
\begin{equation}
    %OK!
    \xi'(Y) = \frac{M_{\mathrm{Pl}}^2 (3 - 4 \alpha \lambda - 2 Y)}{[1 - (3 - 4 \alpha \lambda) Y + Y^2]^2}.
\end{equation}

For the ghost-free candidate $Y_{0, -}$, the second scalar stability condition $\xi'(Y_{0, -}) < 0$ is equivalent to
\begin{equation}
    %OK!
    2Y_{0, -} > 3 - 4 \alpha \lambda.
\end{equation}
This inequality can be expressed in terms of the model parameters as
\begin{equation}
    %OK!
    \lambda > \lambda_{\mathrm{stab}}(\alpha) \coloneq \frac{3 (\alpha + 1) + \sqrt{24 \alpha^2 + 28 \alpha + 9}}{4 \alpha (3 \alpha + 2)}.
\end{equation}
Using the general expression for the scalar sound speed from~\cite{Nishi:2015pta}, we obtain
\begin{equation}
    %OK!
    c_{s, \pm}^2 = \frac{4 \alpha \lambda + 2 Y_{0, \pm} - 3}{A_2 - 10 Y_{0, \pm}}.
\end{equation}
Since
\begin{equation}
    %OK!
    A_2 - 10 Y_{0, \pm} = \mp\sqrt{A_2^2 - 20},
\end{equation}
the squared sound speeds of the two solutions are
\begin{align}
    %OK!
    c_{s,-}^2 &= \frac{4 \alpha \lambda + 2Y_{0, -} - 3}{\sqrt{A_2^2 - 20}}, \\
    c_{s, +}^2 &= -\frac{4 \alpha \lambda + 2Y_{0, +} - 3}{\sqrt{A_2^2 - 20}}.
\end{align}
For the smaller-root solution, the condition $\xi'(Y_{0, -}) < 0$ therefore also ensures $c_{s, -}^2 > 0$.

The tensor stability conditions are again
automatically satisfied for $M_{\mathrm{Pl}}^2 > 0$,
since $g_5 = 0$.

Thus, for $\alpha > 0$ and $\lambda > 0$,
the perturbatively stable Genesis solution
is associated with the smaller root $Y_{0, -}$.
Its parameter region is
\begin{equation}
    %OK!
    \lambda > \lambda_{\mathrm{stab}}(\alpha), \qquad 0 < \alpha \leq 1,
\end{equation}
while for $\alpha > 1$ one additionally has the existence bound
\begin{equation}
    %OK!
    \lambda_{\mathrm{stab}}(\alpha) < \lambda < \frac{9 - 2\sqrt{5}}{4 (\alpha - 1)}.
\end{equation}
To illustrate these results, we choose
\begin{equation}
    %OK!
    \alpha = 2.02, \qquad \lambda = 0.5, \qquad M_{\mathrm{Pl}} = 1.
\end{equation}
For $\alpha = 2.02$, the stability threshold and the two-root existence bound are
\begin{equation}
    %OK!
    \lambda_{\mathrm{stab}} \simeq 0.3355, \qquad \frac{9 - 2 \sqrt{5}}{4 (\alpha - 1)}\simeq 1.1098,
\end{equation}
respectively. Thus, $\lambda = 0.5$ lies within the stable existence interval of the smaller-root solution. As shown below, this choice also gives $0 < c_{s,-}^2 < 1$, while both roots satisfy $\hat{p}(Y_{0, \pm}) < 0$. The value $\alpha = 2.02$ yields the same nearly scale-invariant scalar tilt as in model of~\cref{sec:model1} because the spectral index depends only on $\alpha$ within this class.

For these values, $ A_2 = 6.96$, and the two nontrivial roots are
\begin{align}
    %OK!
    Y_{0, -} &\simeq 0.1627, \\
    Y_{0, +} &\simeq 1.2293.
\end{align}
Both roots satisfy $\hat{p}(Y_{0, \pm}) < 0$ and therefore correspond to expanding Genesis backgrounds.

\begin{figure}
    \centering
    \includegraphics[width=0.8\linewidth]{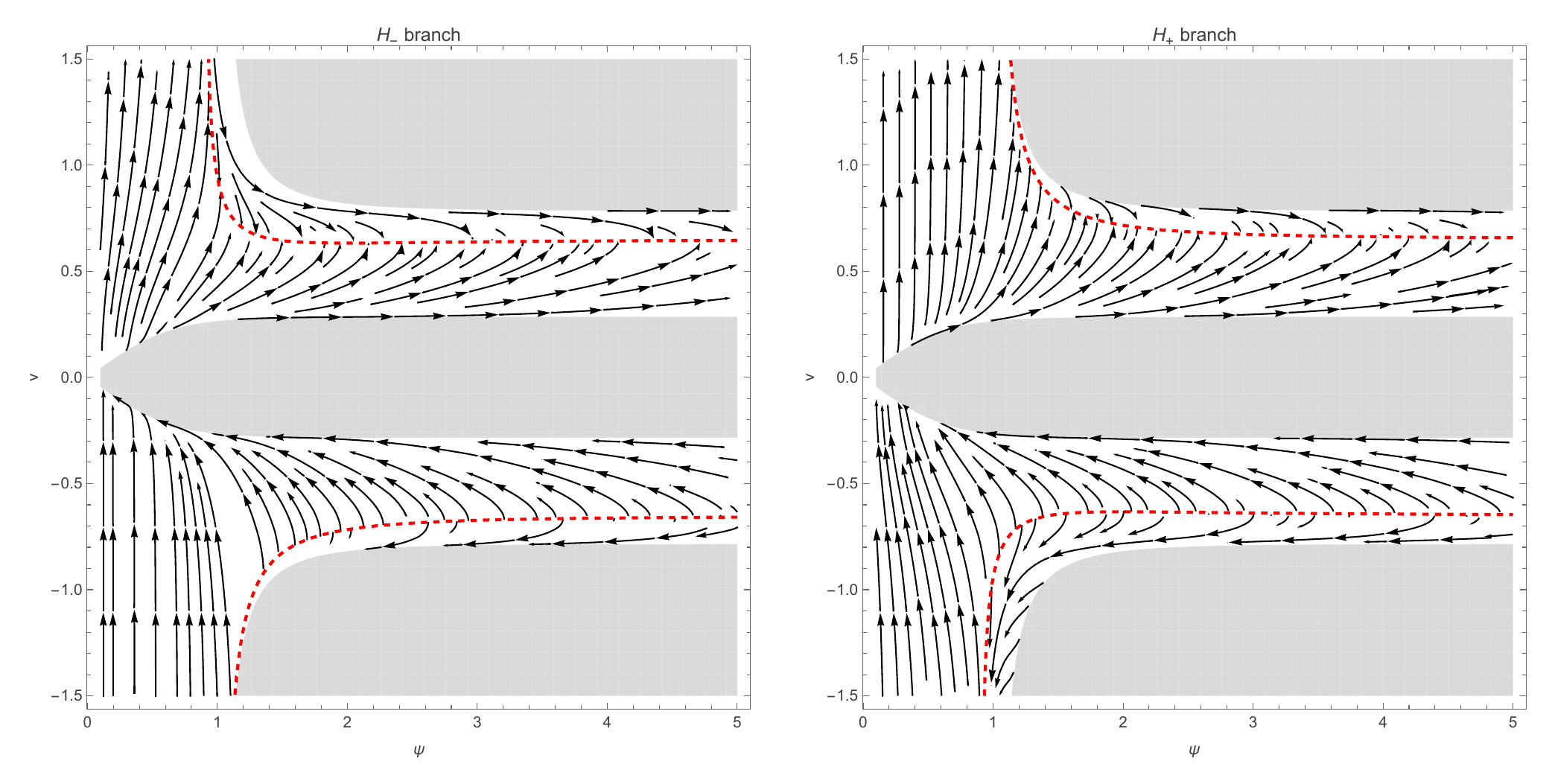}
    \caption{Phase portraits of~\cref{sec:model2} for $\alpha = 2.02$, $\lambda = 0.5$, and $M_{\mathrm{Pl}} = 1$: the $H_-$ branch (left) and the $H_+$ branch (right). Grey shading marks the forbidden region $\mathcal{D} < 0$, where the Friedmann constraint has no real solution for $H$. Arrows indicate increasing cosmic time; in the sector $v < 0$, evolution proceeds toward decreasing $\psi$. The two leading asymptotic Genesis velocities are $v_{\mathrm{G}, -} \simeq -0.2852$ and $v_{\mathrm{G}, +} \simeq -0.7840$. Red dashed curves mark $\Delta = 0$ on the corresponding Friedmann branch, restricted to $v \neq 0$ and $\mathcal{D} > 0$.}
    \label{fig:2}
\end{figure}

The corresponding asymptotic Genesis solutions are
\begin{align}
    %OK!
    v_{\mathrm{G}, -} &\simeq -0.2852, \\
    v_{\mathrm{G}, +} &\simeq -0.7840.
\end{align}
The phase portraits for these parameters are shown in~\cref{fig:2}. The red dashed curves mark the zeros of the determinant $\Delta$ of the coefficient matrix in~\eqref{eq:Reduction_Linear_System}, evaluated on the corresponding Friedmann branch $H_\pm(\psi, v)$ for $v\neq0$ and $\mathcal{D} > 0$. Along these curves, the two equations fail to determine $(\dot{v},\dot{H})$ uniquely, even though the Friedmann constraint admits two distinct real roots.

For the chosen parameters, the squared scalar sound speeds are
\begin{equation}
    %OK!
    c_{s, -}^2 \simeq 0.256, \qquad c_{s, +}^2 \simeq -0.656.
\end{equation}
The smaller-root solution is therefore free from scalar ghost and gradient instabilities in the asymptotic Genesis regime and also satisfies the additional condition $c_{s, -}^2 < 1$. By contrast, the larger-root solution has both a negative scalar kinetic coefficient and $c_{s, +}^2 < 0$.

According to Ref.~\cite{Nishi:2015pta}, the scalar spectral index for $\alpha > 1/2$ is
\begin{equation}
    %OK!
    n_s = 5 - 2\alpha,
\end{equation}
under the matching assumptions discussed in model of~\cref{sec:model1}. For $\alpha = 2.02$, this gives
\begin{equation}
    %OK!
    n_s = 0.96,
\end{equation}
corresponding to a nearly scale-invariant, slightly red-tilted scalar spectrum, see~\cite{Planck:2019nip}.

The tensor sector has the same properties as in model in~\cref{sec:model1}, since $G_4 = M_{\mathrm{Pl}}^2/2$ and $G_5 = 0$. Consequently, tensor perturbations are free from ghost and gradient instabilities and propagate with $c_T^2=1$. Tensor vacuum fluctuations are not amplified in the leading asymptotic Genesis regime; possible tensor production during the subsequent transition requires a separate analysis.

This example demonstrates that two expanding Genesis backgrounds can
coexist in the same theory while possessing different perturbative
properties. For the chosen parameters, the solution associated with
$Y_{0,-}$ is free from linear scalar and tensor ghost and gradient
instabilities in the asymptotic Genesis regime and satisfies
$0<c_{s,-}^2<1$. Under the matching assumption discussed in Model~1,
it also yields a nearly scale-invariant scalar spectrum with
$n_s=0.96$. By contrast, the solution associated with $Y_{0,+}$ has a
negative scalar kinetic coefficient and $c_{s,+}^2<0$. Thus, the
existence of homogeneous Genesis-type attraction does not by itself
imply perturbative viability. As in Model~1, in the subclass with
constant scalar and tensor propagation speeds analyzed in
Ref.~\cite{Akama:2022usl}, avoiding strong coupling while producing a
slightly red-tilted scalar spectrum leads to excessive scalar
non-Gaussianity. The present ``benchmark'' should therefore be
interpreted as a demonstration of the homogeneous phase-space
structure and quadratic perturbative stability, rather than as a
complete perturbatively controlled mechanism for generating the
observed primordial spectrum.

%%%%%%%%%%%%%%%%%%%%%%%%%%%%%%%%%%%%%%%%%%%%%%%%%%%%%%%%%%%%%%%%%%%%%%%
\subsection{Model 3: Power-law contraction preceding a bounce in Horndeski theory}
\label{sec:model3}
%%%%%%%%%%%%%%%%%%%%%%%%%%%%%%%%%%%%%%%%%%%%%%%%%%%%%%%%%%%%%%%%%%%%%%%

We now consider the early contracting stage of the Horndeski bouncing cosmology constructed in~\cite{Ageeva:2022asq}. The covariant Horndeski functions describing this stage are given in Appendix~D of that reference:
\begin{align}
    %OK!
    \label{eq:bounce_G2}
    G_2 &= \hat{g} e^{2 \mu \phi} \left\{e^\phi \left[c_2 e^\phi + \sqrt{2} \left(d_2 - c_3 - 2\mu c_3\right)\sqrt{X}\right] + 2 \mu (d_3 + 2 \mu) X \ln X \right\}, \\
    \label{eq:bounce_G3}
    G_3 &= \frac{\hat{g}}{2} e^{2 \mu \phi}(d_3 + 2 \mu) (2 + \ln X), \\
    \label{eq:bounce_G4}
    G_4 &= \frac{\hat{g}}{2} e^{2 \mu \phi}, \\
    \label{eq:bounce_G5}
    G_5 &= 0.
\end{align}
Here $\hat{g} > 0$, and $c_2, c_3, d_2, d_3, \mu$ are constants. Unlike models of~\cref{sec:model1,sec:model2}, $G_4$ depends on $\phi$; nevertheless, $G_{4 X}=0$ and $G_5 = 0$, so the general reduction scheme developed above remains applicable. Because of the factors $\sqrt{X}$ and $\ln{X}$, the covariant functions are considered on the domain $X > 0$.

Substituting~\cref{eq:bounce_G2,eq:bounce_G3,eq:bounce_G4,eq:bounce_G5} into the homogeneous background equations~\cref{eq:The_Friedmann_Equation,eq:The_Evolution_Equation} give
\begin{equation}
    %OK!
    \label{eq:bounce_energy_phi}
    \mathcal{E} = \hat{g} e^{2 \mu \phi} \left(-3H^2 + 3 d_3 H \dot{\phi} - c_2 e^{2\phi} \right) =0,
\end{equation}
and
\begin{equation}
    %OK!
    \label{eq:bounce_pressure_phi}
    \begin{split}
        \mathcal{P} = \hat{g} e^{2 \mu \phi}\left[c_2e^{2\phi}+3H^2+2\dot H +\sqrt{2} e^\phi \sqrt{X} \bigl(d_2-c_3-2\mu c_3\bigr) \right. \\
        \left. - d_3 (4 \mu X + \ddot{\phi}) + 4 \mu H \dot{\phi} \right] = 0.
    \end{split}
\end{equation}
The logarithmic terms in $G_2$ and $G_3$ cancel in these background equations.

Following Sec.~3.1 of~\cite{Ageeva:2022asq}, we specialize to $\mu > 1$ and $d_3 = -2$. The latter choice is used in that reference to obtain a small scalar sound speed. We employ the field redefinition of~\cref{eq:Reduction_Variables} with $\lambda = 1$, so that $\psi = e^{-\phi}$, $v = \dot{\psi}$, and $X = v^2/(2 \psi^2)$. After dividing~\cref{eq:bounce_energy_phi} by the nonzero factor $-\hat{g} e^{2 \mu \phi}/\psi^2$, the first Friedmann equation becomes
\begin{equation}
    %OK!
    \label{eq:bounce_energy_psi}
    3 \psi^2 H^2 - 6\psi v H + c_2 = 0.
\end{equation}
Its two roots are
\begin{align}
    %OK!
    \label{eq:bounce_H1}
    H_1 &= \frac{v + s(v)}{\psi},\\
    \label{eq:bounce_H2}
    H_2 &= \frac{v - s(v)}{\psi},
\end{align}
where
\begin{equation}
    %OK!
    \label{eq:bounce_s}
    s(v) \coloneq \sqrt{v^2 - \frac{c_2}{3}}.
\end{equation}

The discriminant computed from the original Friedmann constraint~\cref{eq:bounce_energy_phi}, with $d_3=-2$ and $\lambda=1$, is
\begin{equation}
    %OK!
    \label{eq:bounce_discriminant}
    \mathcal{D}(\psi,v)=\frac{12\hat{g}^{\,2}}{\psi^{2+4\mu}}\left(3v^2-c_2\right).
\end{equation}
For $c_2 > 0$, the condition $\mathcal{D}\geq0$ restricts the real phase space to
\begin{equation}
    %OK!
    \label{eq:bounce_allowed_region}
    v^2 \geq \frac{c_2}{3}, \qquad \psi>0.
\end{equation}
Thus the forbidden region is the horizontal strip $\abs{v} < \sqrt{c_2/3}$. Moreover, setting $H = 0$ in \cref{eq:bounce_energy_psi} would require $c_2 = 0$. Hence, for the parameter range considered here, these covariant functions describe the contracting stage rather than the bounce itself. A complete transition to expansion requires the interpolating Lagrangian functions constructed in Appendix~F of Ref.~\cite{Ageeva:2022asq}.

We henceforth restrict to $\mu > 1$, $c_2 > 0$, $\psi > 0$, and the open contracting sector
\begin{equation}
    %OK!
    v < -\sqrt{\frac{c_2}{3}}.
\end{equation}
In this domain, $X > 0$, $s(v) > 0$, and both $H_1$ and $H_2$ are negative. The boundary $s(v) = 0$ is the locus where the two algebraic roots coincide and requires separate treatment using the unreduced background equations. In the chosen sector,
\begin{equation}
    %OK!
    \sqrt{2} e^\phi \sqrt{X} = -\frac{v}{\psi^2}.
\end{equation}
Introducing the combination
\begin{equation}
    %OK!
    \label{eq:bounce_kappa}
    \kappa \coloneq c_3 (1 + 2 \mu) - d_2,
\end{equation}
we obtain the evolution equation
\begin{equation}
    %OK!
    \label{eq:bounce_pressure_psi}
    c_2 + 3 \psi^2 H^2 + 2 \psi^2 \dot{H} + \kappa v + 2 (1 + 2 \mu) v^2 - 2 \psi \dot{v} - 4\mu \psi v H = 0.
\end{equation}
With the original normalization of $\mathcal{E}$ and $\mathcal{P}$, the evolution determinant is
\begin{equation}
    \Delta = \frac{12 \hat{g}^2 v}{\psi^{2 + 4 \mu}}.
\end{equation}
It is strictly negative on both Friedmann branches throughout the contracting sector $v < -\sqrt{c_2/3}$. It also remains nonzero at $s(v) = 0$, where the two Friedmann roots coalesce.

As in the general reduction scheme, differentiating the constraint and eliminating $\dot{H}$ yields $\dot{\psi} = v$ and, on the two algebraic branches,
\begin{align}
    %OK!
    \label{eq:bounce_vdot1}
    \dot v_1 &= -\frac{s(v)}{2 \psi} \left[\kappa + 6 v - 4(\mu - 1) s(v)\right],\\
    \label{eq:bounce_vdot2}
    \dot v_2 &= \frac{s(v)}{2 \psi} \left[\kappa + 6 v + 4(\mu - 1) s(v)\right].
\end{align}
These expressions hold in the sector $v < 0$. For $v > 0$, the $\sqrt{X}$ term has the opposite sign, and the reduced equations must be adjusted accordingly.

The distinguished power-law solutions appear as horizontal trajectories $v = v_* = \mathrm{const}$. For the first and second branches, respectively, their values satisfy
\begin{align}
    %OK!
    \label{eq:bounce_scaling1}
    \kappa + 6 v_*^{(1)} - 4 (\mu - 1) s(v_*^{(1)}) &= 0, \\
    \label{eq:bounce_scaling2}
    \kappa + 6 v_*^{(2)} + 4 (\mu - 1) s(v_*^{(2)}) &= 0.
\end{align}
To connect these trajectories to the contracting backgrounds of.~\cite{Ageeva:2022asq}, recall that
\begin{equation}
    %OK!
    \label{eq:bounce_power_law}
    a(t) = d (-t)^\chi, \qquad H = \frac{\chi}{t}
\end{equation}
for $t < 0$ and $d>0$.
On the exact power-law background, the time coordinate $\hat t$ used in~\cite{Ageeva:2022asq} is related to cosmic time by $t = N\hat t$, where $N > 0$ is constant on that solution. With the time origin chosen consistently, the scalar parametrization gives $\psi = -\hat t = -t/N$. Thus
\begin{equation}
    %OK!
    \label{eq:bounce_vstar_N}
    v_* = -\frac{1}{N}, \qquad H = \frac{\chi v_*}{\psi}.
\end{equation}
The first branch gives
\begin{equation}
    %OK!
    \label{eq:bounce_chi_branch1}
    \chi_1 = 1 + \frac{s(v_*^{(1)})}{v_*^{(1)}}, \qquad 0 < \chi_1 < 1,
\end{equation}
while the second gives $\chi_2 = 1 - s (v_*^{(2)})/v_*^{(2)} > 1$. Hence the first branch reproduces the power-law contraction with $0 < \chi < 1$ studied in~\cite{Ageeva:2022asq}.

The discriminant can also be evaluated explicitly on the power-law background. The background equations imply
\begin{equation}
    %OK!
    c_2 = \frac{3 \chi (2 - \chi)}{N^2}, \qquad N = \frac{2}{\kappa} \left[1 + 2 \mu - 2 (\mu - 1) \chi\right].
\end{equation}
Using $v_* = -1/N$, we therefore obtain
\begin{equation}
    %OK!
    \label{eq:bounce_discriminant_powerlaw}
    \mathcal{D}_{\mathrm{pl}} = \frac{36 \hat{g}^2(1 -\chi)^2}{N^2 \psi^{2 + 4\mu}} = \frac{9 \hat{g}^2 \kappa^2 (1 - \chi)^2}{\left[1 + 2 \mu - 2(\mu - 1) \chi \right]^2 \psi^{2 + 4 \mu}},
\end{equation}
where the subscript ``pl'' means power-law. For $\mu>1$, $\kappa>0$, and $0<\chi<1$, this expression is strictly positive at every finite $\psi>0$. Thus the first-branch power-law solution lies strictly inside the domain admitting two distinct real Friedmann roots. \textit{Within this class, the discriminant condition imposes no additional restriction on the background parameters.}

The analysis of~\cite{Ageeva:2022asq} identifies a useful parameter region in terms of
\begin{equation}
    %OK!
    \label{eq:bounce_rho}
    \varrho \coloneq \frac{c_2}{\kappa^2}.
\end{equation}
For small positive $\mu - 1$, the approximate range is
\begin{equation}
    %OK!
    \label{eq:bounce_original_parameter_region}
    \kappa>0, \qquad 0 < \varrho \lesssim \frac{2}{27}.
\end{equation}
As discussed in Sec.~3.1 of Ref.~\cite{Ageeva:2022asq}, this range
admits the relevant contracting solution with $N>0$ and
$2\mu-3\chi>0$. The latter is the positive-friction condition used in
the perturbation analysis. The distinct condition for decay of the
nonconstant superhorizon mode is
$2\mu+1-3\chi>0$, as discussed below. The power-law background also
satisfies $2\mu>1+\chi$, the condition used to evade the assumptions
of the usual Horndeski no-go result \cite{Kobayashi:2016xpl}. These background and
mode-evolution conditions do not, by themselves, establish the
observational viability or weak-coupling consistency of the model;
the additional restrictions on the tensor-to-scalar ratio and the
strong-coupling scale were studied separately
in~\cite{Ageeva:2022asq}.

For a numerical example, we adopt the parameters used for the contracting stage in Appendix~F of Ref.~\cite{Ageeva:2022asq}:
\begin{equation}
    %OK!
    \label{eq:bounce_parameters}
    \begin{split}
        \hat{g} = 1, \qquad \mu = 1.0099, \qquad c_2 = 0.053, \\
        c_3 = -5, \qquad d_3 = -2, \qquad d_2 = -16.099.
    \end{split}
\end{equation}
They give $\kappa = 1$ and $\varrho = 0.053 < 2/27$. The boundary of the admissible contracting sector is
\begin{equation}
    %OK!
    \label{eq:bounce_vcrit}
    v_{\rm crit} = -\sqrt{\frac{c_2}{3}} \simeq -0.132916.
\end{equation}
Solving \cref{eq:bounce_scaling1,eq:bounce_scaling2} gives
\begin{align}
    %OK!
    v_*^{(1)} &\simeq -0.166010, &\chi_1 &\simeq 0.40087, \\
    v_*^{(2)} &\simeq -0.167338, &\chi_2 &\simeq 1.60753.
\end{align}
In particular, $v_*^{(1)} < v_{\mathrm{crit}}$, so the first power-law solution lies strictly within the admissible contracting region. Using \cref{eq:bounce_vstar_N}, it has $N \simeq 6.02373$, in agreement with the original example. The second solution lies on the other algebraic branch and has $\chi_2 > 1$, outside the contracting class of interest.

\begin{figure}
    \centering
    \includegraphics[width=0.8\linewidth]{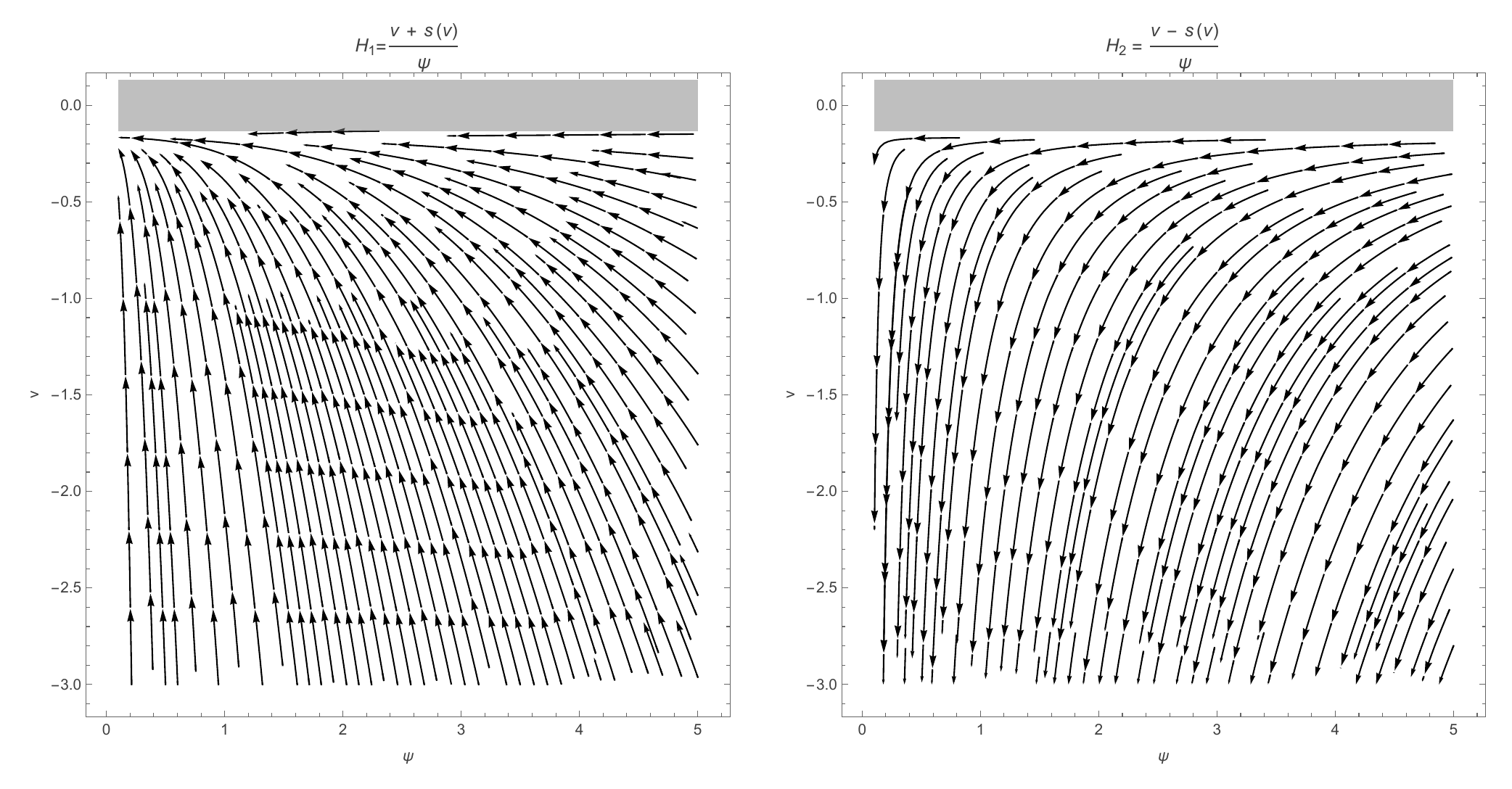}
    \caption{Phase portraits of model given in~\cref{sec:model3} in the $(\psi, v)$ plane for the parameters given in \cref{eq:bounce_parameters}. The left panel corresponds to the branch $H_1 = (v + s(v))/\psi$ and contains the attracting power-law solution $v_*^{(1)} \simeq -0.166010$. The right panel corresponds to $H_2 = (v - s(v))/\psi$ and contains the repelling solution $v_*^{(2)} \simeq -0.167338$. The shaded region represents $\abs{v} < \sqrt{c_2/3}$, where the Friedmann constraint has no real solutions. Both panels illustrate the contracting sector $v < 0$.}
    \label{fig:3}
\end{figure}

To classify the nearby homogeneous trajectories, write the evolution equation on branch $j$ as $\dot{v} = F_j(v)/\psi$ and use $\dot{\psi} = v$. Then
\begin{equation}
    %OK!
    \dv{v}{\psi} = \frac{F_j(v)}{\psi v}.
\end{equation}
For a constant-$v$ solution satisfying $F_j(v_*^{(j)}) = 0$, linearization gives
\begin{equation}
    %OK!
    \dv{\delta v}{\psi} = \frac{\gamma_j}{\psi}\delta v, \quad \gamma_j = \frac{F_j'(v_*^{(j)})}{v_*^{(j)}}.
\end{equation}
Using~\cref{eq:bounce_scaling1,eq:bounce_scaling2}, we obtain
\begin{equation}
    %OK!
    \label{eq:bounce_gamma}
    \delta v \propto \psi^{\gamma_j}, \qquad \gamma_j = 3 (1 - \chi_j) + 2 (\mu - 1).
\end{equation}
For the numerical example,
\begin{equation}
    %OK!
    \gamma_1 \simeq 1.81720, \qquad \gamma_2 \simeq -1.80278.
\end{equation}
Since $\dot{\psi} = v < 0$, increasing cosmic time corresponds to decreasing $\psi$. Consequently, the first power-law solution locally attracts nearby homogeneous trajectories in the chosen phase-space variables, whereas the second repels them. This classification concerns homogeneous background dynamics and is distinct from the stability of inhomogeneous perturbations. The corresponding phase portraits are shown in~\cref{fig:3}.

The two panels of \cref{fig:3} should not be interpreted as
disconnected cosmological histories. At a given point $(\psi,v)$ in
the allowed region, the Friedmann constraint admits two values of the
Hubble parameter. In the contracting sector they obey
$H_2<H_1<0$: both branches therefore describe contraction, with
$H_2$ corresponding to the larger magnitude of the Hubble rate. They
remain distinct for $s(v)>0$ and meet only at
\begin{align}
    v
    &=
    v_{\rm crit}
    =
    -\sqrt{\frac{c_2}{3}},
    &
    H_1
    &=
    H_2
    =
    \frac{v_{\rm crit}}{\psi}
    <0 .
    \label{eq:bounce_fold_location}
\end{align}
This locus is a fold of the projection of the Friedmann-constraint
surface onto the $(\psi,v)$ plane. It is not, by itself, a
discontinuity of the homogeneous geometry; rather, the representation
of the constraint as two separate functions $H_1(\psi,v)$ and
$H_2(\psi,v)$ ceases to be valid there.

A convenient regular coordinate near the fold is
\begin{align}
    y
    &\coloneq
    \psi H-v,
    &
    y^2
    &=
    v^2-\frac{c_2}{3},
    &
    v
    &=
    -\sqrt{\frac{c_2}{3}+y^2},
    \label{eq:bounce_regular_branch_coordinate}
\end{align}
where the last equality selects the component with $v<0$. The branch
$H_2$ corresponds to $y<0$, the branch $H_1$ to $y>0$, and the fold
to $y=0$. Thus, a trajectory that crosses $y=0$ changes its
algebraic branch label without any jump in $\psi$, $v$, or $H$.
Using the unreduced background evolution equation, one finds
\begin{align}
    \dot y
    &=
    -\frac{v}{2\psi}
    \left[
        \kappa+6v-4(\mu-1)y
    \right],
    \nonumber\\
    \left.\dot y\right|_{\rm fold}
    &=
    \frac{\sqrt{c_2/3}}{2\psi}
    \left(
        \kappa-6\sqrt{\frac{c_2}{3}}
    \right).
    \label{eq:bounce_fold_direction}
\end{align}
For the benchmark parameters,
$\kappa-6\sqrt{c_2/3}\simeq0.2025>0$. Hence a homogeneous trajectory
that reaches the fold from $H_2$ continues forward in time onto
$H_1$. Since the common value of $H$ at the fold is negative, this is
a transition between two contracting algebraic sheets, rather than a
reversal of the cosmological evolution.

This possibility does not occur for the two exact power-law
trajectories considered above: both have constant
$v=v_*^{(j)}<v_{\rm crit}$ and therefore remain at
$s(v_*^{(j)})>0$ on their respective branches at every finite
$\psi$. Nor do the nearby trajectories used in
\cref{sec:model3_initial_conditions} cross the fold over the
displayed intervals. In particular,
$v_i^{(2)}=v_*^{(2)}-10^{-4}$ lies on the side of the $H_2$
repeller directed toward more negative $v$, away from
$v_{\rm crit}$. By contrast, for the benchmark parameters, other
$H_2$ initial data satisfying
$v_*^{(2)}<v<v_{\rm crit}$ are directed toward the fold and admit the
homogeneous continuation $H_2\to H_1$. This last statement concerns
the homogeneous background only; perturbative control at the fold is
not established here.

It is useful to compare this result with a related indication of attractor-like behavior found in~\cite{Ageeva:2022asq}. In that work, the nonconstant superhorizon tensor mode was shown to scale as
\begin{equation}
    (-t)^{2 \mu + 1 - 3 \chi},
\end{equation}
and the same condition was found to suppress the corresponding scalar mode. These modes therefore decay during contraction when
\begin{equation}
    2 \mu + 1 - 3 \chi > 0.
\end{equation}
This condition controls the decay of the linear nonconstant scalar
and tensor modes considered in Ref.~\cite{Ageeva:2022asq}; it does not
by itself establish general nonlinear BKL stability. Since
\begin{equation}
    \gamma_j = 3 (1 - \chi_j) + 2 (\mu - 1) = 2 \mu + 1 - 3 \chi_j,
\end{equation}
the homogeneous stability exponent obtained here is governed by precisely the same combination of background parameters. This agreement provides a useful consistency check and shows that the decay condition identified in~\cite{Ageeva:2022asq} is compatible with local attraction of the homogeneous power-law solution. The two results are nevertheless conceptually distinct: \cite{Ageeva:2022asq} concerns long-wavelength scalar and tensor perturbations, whereas the present analysis directly studies neighboring homogeneous background trajectories.

Finally, we recall the perturbative properties of the contracting background. As shown in~\cite{Ageeva:2022asq}, the scalar and tensor spectral tilts satisfy
\begin{equation}
    %OK!
    \label{eq:bounce_spectral_tilts}
    n_s - 1 = n_T = \frac{2 (1 - \mu)}{1 - \chi}.
\end{equation}
Using $\mu=1.0099$ and $\chi=\chi_1\simeq0.400867$, the selected
solution gives
\[
n_s
=
1+\frac{2(1-\mu)}{1-\chi_1}
\simeq 0.967.
\] 
On the selected background with $\chi = \chi_1$, define the dimensionless scalar coefficients by
\begin{equation}
    %OK!
    \mathcal{F}_S = \frac{\hat{g}}{2} \psi^{-2 \mu} f_S, \qquad \mathcal{G}_S = \frac{\hat{g}}{2} \psi^{-2 \mu} g_S.
\end{equation}
For $d_3 = -2$, these coefficients are
\begin{align}
    %OK!
    \label{eq:bounce_scalar_coefficients}
    f_S &= \frac{4 (\mu - 1)}{1 - \chi}, \\
    g_S &= \frac{6}{(1 - \chi)^2},
\end{align}
and
\begin{equation}
    %OK!
    \label{eq:bounce_scalar_sound_speed}
    c_s^2 = \frac{f_S}{g_S} = \frac{2}{3} (\mu - 1) (1 - \chi) \simeq 0.00395.
\end{equation}
The conditions $f_S > 0$ and $g_S > 0$ show the absence of scalar ghost and gradient instabilities on this asymptotic background, and $c_s^2 < 1$ for the chosen parameters. In the tensor sector, $\mathcal{G}_T = \mathcal{F}_T = 2G_4 = \hat{g} \psi^{-2 \mu} > 0$, so tensor perturbations propagate with unit speed. Both tensor coefficients tend to zero as $\psi \to \infty$ in the remote past, realizing the strong gravity in the past behavior of the original model.

For the selected power-law background, the tensor-to-scalar ratio is
\begin{equation}
    %OK!
    r = 8 g_S (c_s^2)^\nu, \qquad \nu = \frac{1 + 2 \mu - 3 \chi}{2 (1 -\chi)}.
\end{equation}
For the chosen parameters, this gives $r \simeq 0.03035$, consistent with the observational bound \cite{Tristram:2020wbi} adopted in~\cite{Ageeva:2022asq}. The primordial non-Gaussianities~\cite{Maldacena:2002vr} of the same contracting model were investigated in~\cite{Ageeva:2024knc}. That analysis uses the conformal transformation to the Einstein frame, where the selected background corresponds to power-law inflation and permits an expansion in slowly varying parameters. Keeping the first nonvanishing contributions gives
\begin{align}
    %OK!
    f_{\mathrm{NL}}^{\mathrm{local}} &\simeq \frac{5 (\mu - 1)}{6 (\mu -\chi)}, \\
    f_{\mathrm{NL}}^{\mathrm{equil}} &\simeq \frac{85}{324} - \frac{5}{81 c_s^2}, \\
    f_{\mathrm{NL}}^{\mathrm{enfold}} &\simeq \frac{1}{32} \left(1 - \frac{1}{c_s^2}\right).
\end{align}
For our parameter choice, these expressions yield
\begin{equation}
    %OK!
    f_{\mathrm{NL}}^{\mathrm{local}} \simeq 0.014, \qquad f_{\mathrm{NL}}^{\mathrm{equil}} \simeq -15.35, \qquad f_{\mathrm{NL}}^{\mathrm{enfold}} \simeq -7.87.
\end{equation}
All three values satisfy the observational bounds considered in~\cite{Ageeva:2024knc}. Furthermore,~\cite{Ageeva:2024knc} shows that, within the parameter region considered there, the non-Gaussianity bounds do not impose additional restrictions once the constraints on the scalar spectral tilt, tensor-to-scalar ratio, and perturbative unitarity are taken into account.

These perturbative constraints are distinct from the homogeneous attractor analysis developed here. For constant $\mu > 1$, extending this power-law model to arbitrarily early times encounters a strong-coupling obstruction: the tensor-sector interaction scale scales as $E_{\mathrm{strong}, T} \propto (-t)^{-\mu}$, whereas the background evolution scale is $E_{\mathrm{class}} \sim \abs{t}^{-1}$. Consequently,
\begin{equation}
    %OK!
    \frac{E_{\mathrm{strong},T}}{E_{\mathrm{class}}} \propto(-t)^{1 - \mu} \longrightarrow 0 \qquad (t \to -\infty).
\end{equation}
A physically controlled application therefore requires a finite interval in which the relevant interaction scales exceed the background and perturbation energy scales, or a modification of the earlier-time evolution~\cite{Ageeva:2022asq}. The present phase portraits classify the classical homogeneous dynamics of the specified contracting model. They do not establish weak coupling throughout the asymptotic past or evolution through a nonsingular bounce; the latter requires the interpolating Lagrangian functions of Ref.~\cite{Ageeva:2022asq}.

%%%%%%%%%%%%%%%%%%%%%%%%%%%%%%%%%%%%%%%%%%%
\section{Sensitivity to initial conditions}
%%%%%%%%%%%%%%%%%%%%%%%%%%%%%%%%%%%%%%%%%%%
\label{sec:Initial_Conditions}

Cosmological attractor behavior has two complementary consequences. In the forward evolution, it suppresses the dependence of the background on small variations of the initial data and can therefore make the subsequent evolution more robust. At the same time, when initially distinct trajectories approach one another, their later states become increasingly difficult to distinguish. Consequently, a small uncertainty in a final quantity may correspond to a much larger uncertainty in the initial value inferred from it. The inverse problem can therefore become poorly conditioned even though the deterministic evolution remains regular and locally unique.

We quantify this effect by comparing homogeneous trajectories at two fixed values of the scalar-field variable, $\psi = \psi_i$ and $\psi = \psi_f$. In the sector considered below, $v = \dot{\psi} < 0$, so $\psi$ decreases monotonically and can be used to parametrize the evolution. Each regular trajectory then defines a local map from the velocity $v_i$ at $\psi=\psi_i$ to the velocity $v_f$ at $\psi = \psi_f$. The derivative of this map measures the forward sensitivity to the initial velocity, while its inverse measures the amplification of uncertainties in reconstructing $v_i$ from $v_f$. This construction quantifies the loss of distinguishability between nearby initial trajectories in the chosen velocity variable; it does not represent a loss of uniqueness, quantum-information scrambling, or an invariant erasure of information from the full cosmological state.

We quantify the local conditioning of reconstructing the initial scalar velocity $v_i$ from its final value $v_f$. Comparisons are performed between fixed sections $\psi = \psi_i$ and $\psi = \psi_f$, with $\psi_i > \psi_f > 0$ in the sector $v < 0$. All Lagrangian parameters and the continuously followed real root of the Friedmann first equation for $H$ are held fixed. For branch $j$, define
\begin{equation}
    \label{J}
    J_j \coloneq \eval{\pdv{v_f}{v_i}}_{\psi_i, \psi_f, j}, \qquad \mathcal{A}_{\mathrm{inv}, j} \coloneq \frac{1}{\abs{J_j}}.
\end{equation}
To linear order, $\delta v_f = J_j \delta v_i$, so $\mathcal{A}_{\mathrm{inv}, j}$ relates infinitesimal absolute errors in the inverse reconstruction. Values greater than unity indicate amplification of inverse errors, whereas values smaller than unity indicate their suppression.

Writing the reduced evolution as $\dv*{v}{\psi} = g_j(\psi, v)$, the Jacobian~\cref{J} obeys
\begin{equation}
    \label{eq:sensitivity_variational_general}
    \dv{J_j}{\psi} = \partial_v g_j(\psi,v_j(\psi))J_j, \qquad J_j(\psi_i) = 1.
\end{equation}
Consequently,
\begin{equation}
    J_j(\psi_f) = \exp\left[\int_{\psi_i}^{\psi_f} \partial_v g_j(\psi, v_j(\psi)) \dd{\psi} \right]>0,
\end{equation}
on every finite regular segment. The section map therefore remains locally invertible; a large inverse sensitivity factor indicates poor conditioning rather than loss of uniqueness.

The calculation requires $v \neq 0$, $\mathcal{D} > 0$, and $\Delta \neq 0$ along the trajectory. The three excluded cases have the distinct meanings discussed in
Sec.~\ref{sec:The_General_Reduction_Scheme}: $\mathcal D=0$ is a
branch-coalescence locus of the Friedmann equation, $v=0$ is a
scalar turning point at which the differentiated constraint loses
scalar information, and $\Delta=0$ on a rolling branch is an
evolution-matrix, and in the present subclass scalar-kinetic,
degeneracy. None of these cases is covered by the branchwise section
map used below. Reconstruction also presupposes knowledge of the branch, which is not determined by $v_f$ alone.

These factors characterize the selected velocity component and section convention. Under a regular change of variable $u = u(\psi, v)$,
\begin{equation}
    J_u = \frac{u_v(\psi_f, v_f)}{u_v(\psi_i, v_i)} J_v.
\end{equation}
For example, $\dot{\phi} = -v/(\lambda\psi)$ gives
\begin{equation}
    J_{\dot{\phi}}=\frac{\psi_i}{\psi_f} J_v.
\end{equation}
Changing either the phase-space coordinate or the comparison sections generally changes the inferred conditioning. The factors below therefore do not constitute invariant measures of information loss or measures of chaotic forward evolution.

%%%%%%%%%%%%%%%%%%%%%%%%%%%%%%%%%%%%%%%%%%%%%%%%%%
\subsection{Generalized Genesis model \textnumero 1}
\label{sec:model1_initial_conditions}
%%%%%%%%%%%%%%%%%%%%%%%%%%%%%%%%%%%%%%%%%%%%%%%%%%

\paragraph{Neglecting gravity.}
In the scalar-field approximation of~\cite{Nishi:2015pta}, the homogeneous equation takes the form
\begin{equation}
    \dv{t}[\psi^{-p} \hat{\rho}(Y)] = 0, \qquad p \coloneq 2 (\alpha + 1),
\end{equation}
and hence admits the first integral
\begin{equation}
    \label{eq:model1_first_integral}
    \hat{\rho}(Y) = C \psi^p, \qquad Y=\frac{v^2}{2 \lambda^2}.
\end{equation}
This approximation neglects expansion-dependent terms in the scalar equation and does not impose $H = 0$ in the Friedmann constraint. For model of~\cref{sec:model1},
\begin{equation}
    \label{eq:model1_rho_ic}
    \hat\rho(Y)= q Y^2 - Y, \qquad q \coloneq 3 - 4 \lambda (\alpha - 1) > 0, \qquad Y_0 = \frac{1}{q}.
\end{equation}
At fixed sections,
\begin{equation}
    \label{eq:model1_initial_to_final}
    \hat\rho(Y_f) = \hat{\rho}(Y_i) \left(\frac{\psi_f}{\psi_i}\right)^p.
\end{equation}
On the monotonic interval containing $Y_0$, with $Y_i, Y_f > 1/(2q)$, the selected solution is
\begin{equation}
    \label{eq:model1_Yf_exact}
    Y_f = \frac{1 + \sqrt{1 + 4 q \hat{\rho}(Y_i)(\psi_f/\psi_i)^p}}{2q}, \qquad v_f = - \lambda \sqrt{2 Y_f}.
\end{equation}
Differentiation gives
\begin{equation}
    \label{eq:model1_decoupled_J}
    J^{\mathrm{asym}} = \left(\frac{\psi_f}{\psi_i}\right)^p
    \frac{v_i}{v_f} \frac{\hat{\rho}'(Y_i)}{\hat{\rho}'(Y_f)}, \qquad \hat\rho'(Y)= 2 q Y - 1,
\end{equation}
where ``asym'' means asymptotic.
This expression is exact within the scalar-field approximation. Near the simple Genesis root, its linearized limit is
\begin{equation}
    \label{eq:model1_decoupled_amp}
    J^{\mathrm{asym, lin}} = \left(\frac{\psi_f}{\psi_i}\right)^p, \qquad \mathcal{A}_{\mathrm{inv}}^{\mathrm{asym, lin}} = \left(\frac{\psi_i}{\psi_f}\right)^p.
\end{equation}
For $\alpha=2.02$, the exponent is $p = 6.04$. Thus the selected velocity component contracts in forward evolution within this approximation.

\paragraph{Including gravitational backreaction.}
For the complete background system,
\begin{equation}
    \label{eq:model1_full_g_ic}
    \dv{v}{\psi} = g_\pm(\psi, v) \coloneq \frac{F(\psi, v, H_\pm(\psi, v))}{v}.
\end{equation}
The Jacobian satisfies
\begin{equation}
    \label{eq:model1_full_j_ic}
    \dv{J_\pm}{\psi} = \partial_v g_\pm(\psi, v_\pm(\psi)) J_\pm, \qquad J_\pm(\psi_i) =1.
\end{equation}
The derivative includes the velocity dependence of the selected Friedmann root:
\begin{equation}
    \partial_v g_\pm = \frac{F_v + F_H H_{\pm,v}}{v} - \frac{F}{v^2}, \qquad H_{\pm,v} = -\frac{\mathcal{E}_v}{\mathcal{E}_H}.
\end{equation}
Here $F_v$ is evaluated at fixed $\psi$ and $H$, and all quantities are subsequently evaluated on the selected branch. The inverse sensitivity factor is
\begin{equation}
    \label{eq:model1_full_amp_ic}
    \mathcal{A}_{\mathrm{inv}, \pm} = \frac{1}{\abs{J_\pm(\psi_f)}}.
\end{equation}
These expressions require $v \neq 0$, $\mathcal{D} > 0$, and $\Delta \neq 0$ throughout the trajectory segment.
For the parameters of~\cref{fig:1}, we choose
\begin{equation}
    v_i = v_{\mathrm{G}} - 10^{-4}, \qquad v_{\mathrm{G}} = -\lambda \sqrt{\frac{2}{3 - 4 \lambda (\alpha - 1)}} \simeq -0.251.
\end{equation}
Although the physical expanding Genesis asymptotics is continuously
associated with the $H_+$ branch, the finite-offset initial data used
here do not lie on the exact finite-$\psi$ Genesis trajectory. For
$v_i=v_{\mathrm G}-10^{-4}$, the Friedmann constraint gives
\begin{equation}
    H_+<0<H_-
\end{equation}
at both initial sections used below, $\psi_i=100$ and $\psi_i=10$.
Thus, the $H_+$ reference trajectories begin in contraction, whereas
the $H_-$ root is an expanding companion algebraic solution rather
than a second realization of the Genesis asymptotics. The two
full-system columns therefore compare branchwise finite-section
sensitivities near the leading asymptotic Genesis velocity, not two
expanding Genesis histories.
The displacement is absolute in the chosen $v=\dot{\psi}$ coordinate,
with $M_{\mathrm{Pl}}=1$. The entries below are obtained by integrating
the reference trajectory together with the variational
equation~\cref{eq:model1_full_j_ic}, using the numerical approach
specified above. As an independent convergence check, we also use the
centered finite-difference estimate
\begin{equation}
    J_\pm^{\mathrm{FD}}(\psi_f)
    \simeq
    \frac{
        v_\pm(\psi_f;v_i+\varepsilon)
        -
        v_\pm(\psi_f;v_i-\varepsilon)
    }{2\varepsilon},
\end{equation}
where $v_\pm(\psi_f;v_i)$ denotes the final velocity on the selected
branch.
This gives
\begin{equation}
    \label{eq:model1_inverse_numerics}
    \begin{array}{c|ccc}
        (\psi_i,\psi_f) & \mathcal A_{\mathrm{inv},-} & \mathcal A_{\mathrm{inv},+} & \mathcal A_{\mathrm{inv},0} \\
        \hline
        (100,50) & 69.24 & 62.45 & 65.80 \\
        (100, 40) & 301.71 & 217.46 & 253.25 \\
        (100, 30) & 4741.29 & 846.52 & 1439.42 \\
        \hline
        (10, 8) & 5.69 & 2.90 & 3.85 \\
        (10, 7.5) & 11.56 & 3.76 & 5.68 \\
        (10, 7) & 40.17 & 4.82 & 8.62 \\
        (10, 6.9) & 65.44 & 5.06 & 9.40 \\
        (10, 6.8) & 154.89 & 5.30 & 10.27 \\

    \end{array}
\end{equation}
The last column, denoted by $\mathcal{A}_{\mathrm{inv}, 0}$, gives the linearized scalar-field estimate $(\psi_i/\psi_f)^p$ from \cref{eq:model1_decoupled_amp}; the subscript $0$ labels this approximation and does not denote a real root of the Friedmann equation for $H$ or the condition $H = 0$. The first two columns give approximate numerical estimates for the full background dynamics on the two Friedmann branches. The comparison shows that gravitational backreaction can substantially modify the inverse sensitivity of the velocity component. 

Throughout the analysis, $\mathcal D>0$ only ensures that the Friedmann roots are real; regular reduced evolution also requires $\Delta\neq0$, and model №2 (Sec.\ref{sec:model2}) contains points with $\mathcal D>0$ but $\Delta=0$. For the model №2 sensitivity data, one initially has $H_+<0<H_-$, so the $H_+$ trajectories begin in contraction. In model №3 (Sec.\ref{sec:model3}), $2\mu-3\chi>0$ is the positive-friction condition, whereas $2\mu+1-3\chi>0$ governs decay of the nonconstant mode; the latter supports suppression only of the linear long-wavelength modes and small anisotropies considered here, not general nonlinear BKL stability. Finally, the repulsion of the second power-law solution is specific to the benchmark parameters in \cref{eq:bounce_parameters} and is not universal for all $\mu>1$.

The rapid increase of $\mathcal{A}_{\mathrm{inv}, -}$ also reflects the geometry of the Friedmann constraint. Near a branch-coalescence locus, $H_{\pm, v} = -\mathcal{E}_v/\mathcal{E}_H$ can become large and strongly affect the projected velocity flow. A dimensionless diagnostic is
\begin{equation}
    \epsilon_{\mathcal D} \coloneq \frac{\sqrt{\mathcal{D}}}{\abs{B} + \sqrt{\mathcal{D}}},
\end{equation}
where $B$ is the linear coefficient of the quadratic Friedmann constraint. For the $H_-$ trajectory from $\psi_i = 10$ to $\psi_f = 6.8$, this quantity decreases from approximately $0.721$ to $0.024$. The large inverse factor should therefore not be attributed solely to attraction toward the Genesis asymptotics.

The distinction between projected and complete-state sensitivity is particularly clear for $(\psi_i, \psi_f) = (100,30)$. Although $J_v \simeq 2.109 \times 10^{-4}$ on $H_-$, the induced Hubble-to-Hubble Jacobian is
\begin{equation}
    \pdv{H_f}{H_i} = \frac{H_{-, v}(\psi_f, v_f)}{H_{-, v}(\psi_i, v_i)} J_v \simeq 0.99999,
\end{equation}
with the initial and final sections held fixed. Thus strong suppression in the $v$-component does not imply comparable suppression in every background variable.

%%%%%%%%%%%%%%%%%%%%%%%%%%%%%%%%%%%%%%%%%%%%%%%%%%
\subsection{Generalized Genesis model №2}
\label{sec:model2_initial_conditions}
%%%%%%%%%%%%%%%%%%%%%%%%%%%%%%%%%%%%%%%%%%%%%%%%%%

\paragraph{Neglecting gravitational backreaction.}
For model of~\cref{sec:model2},
\begin{equation}
    \hat{\rho}(Y) = -Y + A_2 Y^2 - 5Y^3, \qquad \hat{\rho}'(Y) = - 1 + 2 A_2 Y - 15Y^2,
\end{equation}
where $A_2 = 9 - 4 \lambda (\alpha - 1) > 2 \sqrt{5}$. The same first integral \cref{eq:model1_first_integral} and Jacobian formula~\cref{eq:model1_decoupled_J} apply. Because the equation for $Y_f$ is cubic, the final root is selected by continuous continuation from the chosen Genesis root $Y_{0, -}$ or $Y_{0, +}$, remaining on the same local monotonic interval of $\hat\rho(Y)$. The final velocity is $v_f = -\lambda\sqrt{2 Y_f}$.

Both Genesis roots are simple, so linearization around either one gives
\begin{equation}
    \mathcal A_{\mathrm{inv}}^{\mathrm{asym,lin}} = \left(\frac{\psi_i}{\psi_f}\right)^p, \qquad p = 2 (\alpha + 1) = 6.04.
\end{equation}
The common exponent is independent of the sign of $\hat\rho'(Y_0)$. Consequently, the same suppression of velocity differences can occur near both the ghost-free and the ghost-unstable root.

\paragraph{Including gravitational backreaction.}
For the full background system, we integrate each reference trajectory
together with the corresponding variational equation on each
Friedmann branch, using the same numerical method as in Model~1.
Centered finite differences are used only as an independent
convergence check.
For the parameters of Model~2, $\alpha = 2.02$, $\lambda  =0.5$ and $M_{\mathrm{Pl}}=1$, we find $v_{\mathrm G, -} \simeq -0.2852$ and $v_{\mathrm G, +} \simeq -0.7840$. The initial data are chosen at $\psi_i = 100$ as $v_i = v_{\mathrm{G}, -} - 10^{-4}$ and $v_i = v_{\mathrm G, +} + 10^{-4}$, respectively, so that both initial velocities lie between the Genesis values. At $\psi_f=50$, the inverse sensitivity factors are
\begin{equation}
    \begin{array}{c|ccc}
         & \mathcal{A}_{\mathrm{inv}, -}
         & \mathcal{A}_{\mathrm{inv}, +}
         & \mathcal{A}_{\mathrm{inv}}^{\mathrm{asym,exact}} \\ \hline
        Y_{0, -} & 66.55 & 64.87 & 65.70 \\
        Y_{0, +} & 77.39 & 57.35 & 65.88
    \end{array}
\end{equation}
The last column uses the nonlinear first-integral relation and its Jacobian for the same initial velocities. The full results depend on both the selected Friedmann branch and the Genesis root. Both roots exhibit forward contraction of the velocity component over the chosen interval, although the larger-root background is perturbatively unstable. This comparison explicitly separates homogeneous attraction from perturbative viability.

%%%%%%%%%%%%%%%%%%%%%%%%%%%%%%%%%%%%%%%%%%%%%%%%%%
\subsection{Contracting stage of the Horndeski Bounce}
\label{sec:model3_initial_conditions}
%%%%%%%%%%%%%%%%%%%%%%%%%%%%%%%%%%%%%%%%%%%%%%%%%%

Unlike models of Genesis, current model describes power-law contraction. Using the notation $\dot{v} = F_j(v)/\psi$ introduced above, its reduced equations become
\begin{equation}
    \label{eq:model3_ic_autonomous}
    \dv{v}{\ln\psi} = Q_j(v), \quad Q_j(v) \coloneq \frac{F_j(v)}{v}, \quad j = 1, 2,
\end{equation}
where
\begin{align}
    Q_1(v) &= -\frac{s(v)}{2v} \left[\kappa + 6 v - 4 (\mu - 1) s(v)\right], \\
    Q_2(v) &= \frac{s(v)}{2v} \left[\kappa + 6 v + 4 (\mu - 1) s(v)\right].
\end{align}
For a nonstationary regular trajectory on which $Q_j \neq 0$,
\begin{equation}
    \int_{v_i}^{v_f}\frac{\dd{u}}{Q_j(u)} = \ln\frac{\psi_f}{\psi_i}.
\end{equation}
Differentiation at fixed sections gives
\begin{equation}
    -\frac{1}{Q_j(v_i)} + \frac{J_j}{Q_j(v_f)} = 0,
\end{equation}
and therefore
\begin{equation}
    \label{eq:model3_ic_exact}
    J_j = \frac{Q_j(v_f)}{Q_j(v_i)}, \qquad \mathcal{A}_{\mathrm{inv}, j} = \abs{\frac{Q_j(v_i)}{Q_j(v_f)}}.
\end{equation}

The constant-velocity trajectory $Q_j(v_*^{(j)}) = 0$ is regular, although the ratio formula takes the indeterminate form $0/0$ there. Its Jacobian follows directly from the variational equation. Since $Q_j'(v_*^{(j)}) = \gamma_j$,
\begin{equation}
    \label{eq:model3_ic_linear}
    J_j=\left(\frac{\psi_f}{\psi_i}\right)^{\gamma_j}, \quad \mathcal{A}_{\mathrm{inv}, j} = \left(\frac{\psi_i}{\psi_f}\right)^{\gamma_j}, \quad \gamma_j = 3( 1 - \chi_j) + 2(\mu - 1).
\end{equation}
These expressions give the exact infinitesimal sensitivity on the constant-velocity reference trajectory and the leading approximation for nearby reference trajectories. For the benchmark, $\gamma_1 \simeq 1.8172$ and $\gamma_2 \simeq -1.8028$.

Autonomy in $\ln\psi$ implies that, for fixed initial velocity and model parameters, the section map depends on $\psi_f/\psi_i$ rather than on the absolute section values.
For the parameters of \cref{eq:bounce_parameters}, we consider initial velocities $v_i^{(j)} = v_*^{(j)} - 10^{-4}$, $j = 1,2$. The inverse sensitivity factors obtained from the full reduced dynamics are compared with the linearized estimates in
\begin{equation}
    \label{eq:model3_inverse_numerics}
    \begin{array}{c|cccc}
        (\psi_i,\psi_f)
        & \mathcal A_{\mathrm{inv},1}
        & \mathcal A_{\mathrm{inv},2}
        & \mathcal A_{\mathrm{inv},1}^{\mathrm{lin}}
        & \mathcal A_{\mathrm{inv},2}^{\mathrm{lin}}
        \\ \hline
        (100,80) & 1.501 & 0.668 & 1.500 & 0.669 \\
        (100,70) & 1.914 & 0.525 & 1.912 & 0.526 \\
        (100,50) & 3.529 & 0.285 & 3.524 & 0.287 \\
        (100,40) & 5.295 & 0.190 & 5.286 & 0.192 \\
        (100,30) & 8.933 & 0.112 & 8.916 & 0.114 \\
        \hline
        (10,8)  & 1.501 & 0.668 & 1.500 & 0.669 \\
        (10,5)  & 3.529 & 0.285 & 3.524 & 0.287
    \end{array}
\end{equation}
Here the linearized values are calculated using $\mathcal A_{\mathrm{inv}, j}^{\mathrm{lin}} = (\psi_i/\psi_f)^{\gamma_j}$.
The entries in \cref{eq:model3_inverse_numerics} are shown to
three decimal places because the deviations of the full results
from the linearized estimates are at the $10^{-3}$--$10^{-2}$
level in absolute value.

The agreement shows that the chosen trajectories remain close enough to the power-law solutions for the linearized estimates to be accurate over the displayed intervals. Equal factors for section pairs with the same ratio follow from the autonomy of the reduced equations.

On the attracting branch $H_1$, the final velocity is less sensitive to its initial value, so inverse errors are amplified. On the repelling branch $H_2$, the forward velocity dependence grows and inverse errors are suppressed. These statements concern the selected velocity component during pre-bounce contraction.

The subsequent bounce can change the transfer of homogeneous deviations between contraction and expansion. Determining whether the suppression of the velocity component persists requires evolving perturbations of the homogeneous initial data through the full interpolating model. That calculation is beyond the present analysis.

%%%%%%%%%%%%%%%%%%%%%
\section{Conclusions}
\label{sec:Conclusions}
%%%%%%%%%%%%%%%%%%%%%

Building on the phase-space analysis of generalized Galilean Genesis in~\cite{Nishi:2015pta}, we made the corresponding reduction explicit and formulated it in a common form for homogeneous Horndeski models with $G_4 = G_4(\phi)$ and $G_5 = 0$, retaining both real roots of the quadratic background equation for the Hubble parameter. 

For the two Genesis models, the scalar-field approximation gives the same leading suppression law,
\begin{equation*}
    \delta v \propto \psi^{2(\alpha+1)},
\end{equation*}
near any simple nonzero Genesis root.
The sensitivity analysis provides several additional results beyond the qualitative statement that Genesis is an attractor. First, it quantifies how strongly the dependence on the initial scalar velocity is suppressed over a finite interval of evolution. Second, it shows that gravitational backreaction can substantially modify this finite-time suppression and can produce different results for the two real values of the Hubble parameter allowed by the background equation. Third, in the model with two Genesis solutions, the same leading suppression of velocity differences occurs near both the perturbatively healthy solution and the ghost-unstable one. Homogeneous attraction therefore does not imply perturbative viability. Finally, very large inverse sensitivity factors in the velocity variable can be enhanced when the two real roots for the Hubble parameter become close to one another. In this regime, strong suppression of differences in $v$ need not be accompanied by comparable suppression of differences in $H$. \textit{The sensitivity analysis therefore clarifies both how gravity
modifies the finite-time branchwise flow near the leading Genesis
velocities and how strongly the apparent convergence depends on the
variable used to describe the homogeneous state.}

For the power-law contracting model, the reduced dynamics is autonomous in $\ln\psi$ and admits the exact section-map Jacobian
\begin{equation*}
    J_j = \frac{Q_j(v_f)}{Q_j(v_i)}
\end{equation*}
on nonstationary regular trajectories. Near the constant-velocity solutions, the sensitivity is governed by
\begin{equation*}
    \gamma_j = 3 (1 - \chi_j)+2(\mu - 1).
\end{equation*}
For the chosen parameters,
\begin{equation*}
    \gamma_1\simeq1.8172, \qquad \gamma_2\simeq-1.8028.
\end{equation*}
The physically relevant power-law contracting solution therefore attracts nearby homogeneous trajectories, whereas the second power-law solution repels them. The finite-time sensitivity factors also confirm that the linearized power-law approximation remains accurate for the nearby trajectories considered in our numerical examples.

These results quantify the dependence on the initial scalar velocity between fixed values of the scalar-field variable $\psi$ and for a prescribed continuous choice of the real Hubble solution. They provide a finite-time measure of the convergence or divergence of nearby homogeneous trajectories, rather than a direct reconstruction of primordial initial conditions from cosmological observations. The numerical sensitivity depends on the chosen phase-space variable and on the comparison sections. In particular, strong suppression in the velocity component does not necessarily imply comparable suppression in the Hubble parameter.

On every finite regular interval considered here, the map from the initial to the final velocity remains locally invertible. The inverse problem can nevertheless become poorly conditioned, so that small uncertainties in the final velocity correspond to much larger uncertainties in the reconstructed initial value, without any loss of uniqueness. The cosmological interpretation of these results additionally requires perturbative control throughout the relevant interval of evolution.

For the contracting model, a natural next step is to extend the sensitivity analysis through the bounce using the interpolating Lagrangian functions constructed in~\cite{Ageeva:2022asq}. Evolving nearby homogeneous solutions through this transition would determine whether the suppression of the initial-velocity dependence generated during contraction persists, is weakened, or is enhanced during the bounce and the subsequent expanding stage.    

%%%%%%%%%%%%%%%%%%%%%%%%%%
\section{Acknowledgements}
%%%%%%%%%%%%%%%%%%%%%%%%%%

The work of YA on Sections~2 and 3 has been supported by the Russian Science Foundation Grant No.~24-72-10110. AS thanks the Theoretical Physics and Mathematics Advancement Foundation “BASIS” for the student fellowships under the contract 24-2-10-33-1.

The authors thank Sergei Mironov for reading the manuscript and providing useful comments and discussion. AS thanks Andrey Pivovarov for helpful discussions.

The authors used publicly available large language models to assist with English-language editing and selected technical coding tasks, including implementing established algorithms and reviewing source code. The tools were not used for scientific interpretation or for generating the reported results.

\bibliography{citation}

\end{document}